\documentclass[aps, pra, twocolumn,amsmath,amssymb,floatfix, superscriptaddress]{revtex4-1} %preprintnumbers,
\usepackage{graphicx}
\usepackage{physics}
\usepackage{hyperref} % For hyperlinks in the PDF
\usepackage[utf8]{inputenc}
\usepackage[english]{babel}
\hypersetup{colorlinks,allcolors=blue}
\begin{document}

\title{Plane and Stripe Wave Phases of a Spin-Orbit Coupled Bose-Einstein Condensate in an Optical Lattice with a Zeeman Field}

\author{Kristian Mæland}
    \affiliation{\mbox{Center for Quantum Spintronics, Department of Physics, Norwegian University of Science and Technology,}\\NO-7491 Trondheim, Norway} 
\author{Andreas T.G. Janssønn}
    \affiliation{\mbox{Center for Quantum Spintronics, Department of Physics, Norwegian University of Science and Technology,}\\NO-7491 Trondheim, Norway} 
\author{Jonas H. Rygh}
    \affiliation{Department of Physics, Norwegian University of Science and Technology, NO-7491 Trondheim, Norway}
\author{Asle Sudbø}
\email[Corresponding author: ]{asle.sudbo@ntnu.no}
\affiliation{\mbox{Center for Quantum Spintronics, Department of Physics, Norwegian University of Science and Technology,}\\NO-7491 Trondheim, Norway}

\date{\today} 

\begin{abstract}
A weakly interacting, spin-orbit coupled, ultracold, dilute Bose gas on a two-dimensional square lattice with an external Zeeman field is studied. We explore the plane and stripe wave phases of the system involving nonzero condensate momenta, which occur when the Zeeman field is below a critical value. Their excitation spectra are found using Bogoliubov theory and by two different routes. The validity of each method to obtain the excitation spectrum is discussed, and it is found that projection on the lowest single-particle band is an excellent approximation in the plane wave phase, while it is a poor approximation in the stripe wave phase. While the plane wave phase has a phonon minimum at its single condensate momentum, revealing a nonzero sound velocity of the excitations, the stripe wave phase has quadratic minima at its two condensate momenta showing zero sound velocity of the excitations. We discuss how the presence of more than one condensate momentum is essential for these differences between the two phases. Additionally, it is emphasized that the zero sound velocity in the stripe wave phase is a lattice effect, since continuum studies of the same phase have shown nonzero sound velocity.
\end{abstract}

\maketitle

%------------------------------------------INTRODUCTION--------------------------------------------------------------------------
\section{Introduction}
The first experimental achievement of Bose-Einstein condensation (BEC) in ultracold, dilute atomic gases \cite{Cornell} sparked a flurry of activity in the physics of cold atoms, and great strides have since been made in manufacturing and manipulating such condensates. The studies were extended to include the effect of spin-orbit coupling (SOC) when a synthetic SOC was demonstrated experimentally \cite{lin2011expSOC}. Due to the Doppler effect, lasers can induce momentum dependent transitions between two pseudospin states, emulating the SOC of spin-1/2 particles. The methods have since been refined, and highly tunable synthetic SOC with different linear combinations of Rashba \cite{RashbaSOC} and Dresselhaus \cite{DresselhausSOC} SOC have been achieved experimentally \cite{spielmantunable, wu2016realization2DSOC, galitskispielmanSOCrev, spielman2016rashba, wang2DRashba, zhaiSWRashbaRev}. The introduction of SOC to the ultracold gas has many interesting consequences including the lack of Galilean invariance \cite{HamnerSOCGalileanExp} and hence a frame dependent superfluid velocity \cite{37}. This greatly complicates the theoretical treatment of such condensates. 

It is also possible to load the atoms onto an optical lattice, since lasers can generate a periodic potential landscape \cite{PethickSmith}. In that case, the highly tunable experimental setup can be used to simulate numerous condensed matter physics phenomena under completely controlled conditions. Examples where SOC plays an important role are the quantum spin Hall effect and topological insulators \cite{manchon2015newRashbaApp}. Furthermore, the controllability of atoms trapped in optical lattices means they could find applications in quantum computing \cite{Jaksch}.

In this paper, we consider a two-dimensional (2D), Rashba SOC, weakly interacting BEC in the presence of a square optical lattice and an external Zeeman field. An important consequence of the SOC is the presence of phases with nonzero condensate momenta, some of which can be viewed as bosonic analogs of Fulde-Ferrell-Larkin-Ovchinnikov states in superconductors \cite{FF, LO, FFLO}. The Fulde-Ferrell analogous plane wave (PW) phase with one nonzero condensate momentum was treated in \cite{Toniolo} by projection on the lowest single particle band. In this paper we will further explore the Larkin-Ovchinnikov analogous stripe wave (SW) phase with two oppositely directed condensate momenta. This phase has previously been studied in a continuum \cite{li2013superstripes, Swexcitation3Dcont}, and was later observed experimentally \cite{SWdetectedExp}, but its excitation spectrum has to our knowledge not been obtained in the presence of a lattice. The excitation spectra in the two phases are found by the same method used in \cite{Toniolo} projecting down on the lowest energy excitations, as well as without any projection.  
It is found that the excitations in the SW phase have zero sound velocity, unlike the nonzero sound velocity found in \cite{Toniolo} for the PW phase and in \cite{li2013superstripes, Swexcitation3Dcont} for the SW phase in a continuum. In addition, the method used in \cite{Toniolo} is found to be an excellent approximation in the PW phase, while it fails at almost all parameters in the SW phase. The origin of these results will be discussed.

%----------------------------------------------------BOGOLIUBOV THEORY-----------------------------------------------------------------
\section{Bogoliubov Theory} \label{sec:develop}
We start with a Bose-Hubbard Hamiltonian for a Bose gas with two spin components akin to that introduced in \cite{LS}, and include a Rashba SOC discretized to a lattice formulation
\begin{align}
    \begin{split}
    \label{eq:realspaceBoseHubbard}
        H &= -\sum_\alpha t^\alpha\sum_{\langle i, j \rangle}  b_i^{\alpha\dagger}b_j^\alpha - \sum_\alpha \mu^\alpha \sum_i  b_i^{\alpha\dagger}b_i^\alpha \\
        &-i\lambda_R \sum_{\alpha\beta}\sum_{i,n}\Big(b_i^{\alpha\dagger}\hat{z}\cdot( \boldsymbol{\sigma}^{\alpha\beta}\cross \hat{\boldsymbol{a}}_n) b_{i+n}^\beta - \textrm{H.c.} \Big)\\
        &+\frac{1}{2}\sum_{\alpha\beta} U^{\alpha\beta} \sum_{i}  b_i^{\alpha\dagger}b_i^{\beta\dagger}b_i^\beta b_i^\alpha.
    \end{split}
\end{align}
Here, $b_i^{\alpha}$ annihilates a boson of spin $\alpha$ at the lattice site $i$, $t^\alpha$ is a spin dependent nearest neighbor hopping parameter and $\mu^\alpha$ is a spin dependent chemical potential. $\lambda_R$ is the strength of the Rashba SOC, $\boldsymbol{\sigma}$ is a vector containing the Pauli matrices, $\boldsymbol{a}_n$ are the $d$ primitive vectors of a $d$-dimensional Bravais lattice, and hats denote unit vectors. The two spin components are spin up and spin down, while H.c. indicates the Hermitian conjugate of the preceding term. The interactions are assumed to be repulsive and $U^{\alpha\beta}$ is the interaction parameter for an on-site two-body scattering involving particles with spin $\alpha$ and $\beta$.
The main candidates for experimental realization of this Hamiltonian are ultracold gases of bosonic atoms, where external lasers set up an optical lattice and a synthetic SOC. Experimentalists choose two hyperfine states of the atoms as the two components of the system, which are then labeled pseudospin up and pseudospin down.

It is advantageous to consider the system in momentum space, since BEC is associated with the particles' momentum distribution. The bosonic operators are Fourier transformed using $b_i^\alpha = (1/\sqrt{N_s})\sum_{\boldsymbol{k}} A_{\boldsymbol{k}}^\alpha e^{-i \boldsymbol{k}\cdot \boldsymbol{r}_i}$, where $N_s$ is the number of lattice sites, $A_{\boldsymbol{k}}^\alpha$ is a bosonic operator annihilating a boson with spin $\alpha$ and momentum $\boldsymbol{k}$, and $\boldsymbol{r}_i$ is the position of lattice site $i$. 
Technically, $\boldsymbol{k}$ is a quasimomentum limited to the first Brillouin zone (1BZ). The system is periodic in terms of $\boldsymbol{k}$ by the size of the 1BZ, and for the remainder of the paper we will use the name momentum in place of quasimomentum.
In momentum space, the Hamiltonian becomes
\begin{align}
    \begin{split}
    \label{eq:FullH}
        H &= \sum_{\boldsymbol{k}}\sum_{\alpha\beta}\eta_{\boldsymbol{k}}^{\alpha\beta}A_{\boldsymbol{k}}^{\alpha\dagger}A_{\boldsymbol{k}}^\beta   \\
        &+\frac{1}{2N_s}\sum_{\boldsymbol{k}\boldsymbol{k}'\boldsymbol{p}\boldsymbol{p}'}\sum_{\alpha\beta}U^{\alpha\beta} A_{\boldsymbol{k}}^{\alpha\dagger}A_{\boldsymbol{k}'}^{\beta\dagger}A_{\boldsymbol{p}}^{\beta}A_{\boldsymbol{p}'}^{\alpha} \delta_{\boldsymbol{k}+\boldsymbol{k}', \boldsymbol{p}+\boldsymbol{p}'},
    \end{split}
\end{align}
where
\begin{gather}
    \eta_{\boldsymbol{k}} = \begin{pmatrix} \epsilon_{\boldsymbol{k}}^{\uparrow} -\mu^{\uparrow} & s_{\boldsymbol{k}} \\ s_{\boldsymbol{k}}^* &  \epsilon_{\boldsymbol{k}}^{\downarrow} -\mu^{\downarrow} \end{pmatrix},
\end{gather}
\begin{equation}
\label{eq:generalek}
    \epsilon_{\boldsymbol{k}}^\alpha \equiv -2t^\alpha \sum_{n=1}^d \cos(\boldsymbol{k}\cdot \boldsymbol{a}_n)
\end{equation}
and the Rashba SOC term is
\begin{equation}
    \label{eq:Bravaissk}
    s_{\boldsymbol{k}} \equiv -2\lambda_R \sum_{n=1}^d (\hat{\boldsymbol{a}}_n \cdot \hat{\boldsymbol{y}} + i \hat{\boldsymbol{a}}_n \cdot \hat{\boldsymbol{x}} )\sin(\boldsymbol{k} \cdot \boldsymbol{a}_n).
\end{equation}

\subsection{Mean-Field Theory}
We assume the temperature is low enough for BEC to occur, such that the condensate is dominant. We will set the temperature to zero in the calculations, and consider quantum fluctuations of the ground state. It is assumed that there are few excitations, and terms in the Hamiltonian involving a product of three or more excitation operators are therefore neglected. $A_{\boldsymbol{k}_{0i}}^\alpha$ is named a condensate operator if $\boldsymbol{k}_{0i}$ is any occupied condensate momentum, while $A_{\boldsymbol{k}}^\alpha$ is an excitation operator given that $\boldsymbol{k} \neq \boldsymbol{k}_{0i}$. The momentum configurations in the interaction terms that include at most two excitation momenta are represented in table \ref{tab:interactionmomentumconfig}. The cases 2-5 lead to terms that are linear in excitation operators, and originate from the fact that the momentum conservation may be obeyed by three condensate momenta and one excitation momentum. This possibility, requiring multiple condensate momenta in the system, was first elucidated by Janssønn \cite{master} and has to our knowledge not been explored previously. Inserting these momentum configurations yields
\begin{equation}
\label{eq:MFTH}
    H \approx H_0 + H_1 + H_2,
\end{equation}
where
\begin{align}
\label{eq:H0}
    \begin{split}
        H_0 = &\sum_{i}  \sum_{\alpha\beta}\eta_{\boldsymbol{k}_{0i}}^{\alpha\beta}A_{\boldsymbol{k}_{0i}}^{\alpha\dagger}A_{\boldsymbol{k}_{0i}}^\beta \\
        &+ \frac{1}{2N_s}\sum_{iji'j'}\sum_{\alpha\beta}U^{\alpha\beta}A_{\boldsymbol{k}_{0i}}^{\alpha\dagger}A_{\boldsymbol{k}_{0j}}^{\beta\dagger}A_{\boldsymbol{k}_{0i'}}^\beta A_{\boldsymbol{k}_{0j'}}^\alpha\\
        &\mbox{\qquad\qquad\qquad}\cdot \delta_{\boldsymbol{k}_{0i}+\boldsymbol{k}_{0j},\boldsymbol{k}_{0i'}+\boldsymbol{k}_{0j'}},
    \end{split}
\end{align}
\begin{align}
\label{eq:H1}
    \begin{split}
        H_1 &= \frac{1}{N_s}\left.\sum_{\boldsymbol{k}}\right.^{'}\sum_{iji'}\sum_{\alpha\beta}U^{\alpha\beta}\big(A_{\boldsymbol{k}_{0i}}^{\alpha\dagger}A_{\boldsymbol{k}_{0j}}^{\beta\dagger}A_{\boldsymbol{k}_{0i'}}^\beta A_{\boldsymbol{k}}^\alpha \\
        &\mbox{\qquad\qquad}+A_{\boldsymbol{k}}^{\alpha\dagger}A_{\boldsymbol{k}_{0i'}}^{\beta\dagger}A_{\boldsymbol{k}_{0j}}^\beta A_{\boldsymbol{k}_{0i}}^\alpha \big)\delta_{\boldsymbol{k}+\boldsymbol{k}_{0i'},\boldsymbol{k}_{0i}+\boldsymbol{k}_{0j}}
    \end{split}
\end{align}
and
\begin{align}
    \begin{split}
    \label{eq:H2}
        H_2 = &\left.\sum_{\boldsymbol{k}}\right.^{'}\sum_{\alpha\beta}\eta_{\boldsymbol{k}}^{\alpha\beta}A_{\boldsymbol{k}}^{\alpha\dagger}A_{\boldsymbol{k}}^\beta \\
        &+\frac{1}{2N_s}\left.\sum_{\boldsymbol{k}\boldsymbol{k}'}\right.^{''}\sum_{ij}\sum_{\alpha\beta} U^{\alpha\beta}\Big(\big( A_{\boldsymbol{k}_{0i}}^{\alpha\dagger}A_{\boldsymbol{k}_{0j}}^{\beta\dagger}A_{\boldsymbol{k}}^\beta A_{\boldsymbol{k}'}^\alpha\\
        &\mbox{\qquad\quad}+ A_{\boldsymbol{k}}^{\alpha\dagger}A_{\boldsymbol{k}'}^{\beta\dagger}A_{\boldsymbol{k}_{0j}}^\beta A_{\boldsymbol{k}_{0i}}^\alpha\big)\delta_{\boldsymbol{k}+\boldsymbol{k}',\boldsymbol{k}_{0i}+\boldsymbol{k}_{0j}} \\
        &\mbox{\qquad\quad}+2\big(A_{\boldsymbol{k}_{0i}}^{\alpha\dagger}A_{\boldsymbol{k}}^{\beta\dagger}A_{\boldsymbol{k}_{0j}}^\beta A_{\boldsymbol{k}'}^\alpha \\
        &\mbox{\qquad\quad}+A_{\boldsymbol{k}_{0i}}^{\alpha\dagger}A_{\boldsymbol{k}}^{\beta\dagger}A_{\boldsymbol{k}'}^\beta A_{\boldsymbol{k}_{0j}}^\alpha \big)\delta_{\boldsymbol{k}+\boldsymbol{k}_{0i},\boldsymbol{k}'+\boldsymbol{k}_{0j}}\Big).
    \end{split}
\end{align}
The primes indicate that the sums exclude any occupied condensate momenta.

\begin{table}[ht]
    \centering
    \caption{The momentum configurations in the interaction terms with at most two excitation momenta. Table reproduced from \cite{master}.}
    \begin{tabular}{|c|c|c|c|c|c|c|c|c|c|c|c|}
        \hline
        Case & 1 &2 & 3 &4 & 5 &6 &7 &  8 &  9 &10 & 11 \\ 
        \hline
        $\boldsymbol{k}$ & $\boldsymbol{k}_{0i}$ &$\boldsymbol{k}_{0i}$ &$\boldsymbol{k}_{0i}$ &$\boldsymbol{k}_{0i}$ &$\boldsymbol{k}$ & $\boldsymbol{k}_{0i}$ & $\boldsymbol{k}_{0i}$ & $\boldsymbol{k}_{0i}$ &  $\boldsymbol{k}$ &$\boldsymbol{k}$ &  $\boldsymbol{k}$ \\
        \hline
        $\boldsymbol{k}'$ & $\boldsymbol{k}_{0j}$ &$\boldsymbol{k}_{0j}$ &$\boldsymbol{k}_{0j}$ &$\boldsymbol{k}'$ & $\boldsymbol{k}_{0j}$ & $\boldsymbol{k}_{0j}$ & $\boldsymbol{k}'$ & $\boldsymbol{k}'$ &  $\boldsymbol{k}'$ & $\boldsymbol{k}_{0j}$ &$\boldsymbol{k}_{0j}$ \\ 
        \hline
        $\boldsymbol{p}$ & $\boldsymbol{k}_{0i'}$ &$\boldsymbol{k}_{0i'}$ & $\boldsymbol{p}$ & $\boldsymbol{k}_{0i'}$ &  $\boldsymbol{k}_{0i'}$ &  $\boldsymbol{p}$ &  $\boldsymbol{k}_{0i'}$ &$\boldsymbol{p}$ &$\boldsymbol{k}_{0i'}$ & $\boldsymbol{p}$ & $\boldsymbol{k}_{0i'}$ \\
        \hline
        $\boldsymbol{p}'$ & $\boldsymbol{k}_{0j'}$ & $\boldsymbol{p}'$ &$\boldsymbol{k}_{0j'}$ & $\boldsymbol{k}_{0j'}$ &  $\boldsymbol{k}_{0j'}$ & $\boldsymbol{p}'$ &  $\boldsymbol{p}'$ &  $\boldsymbol{k}_{0j'}$ &  $\boldsymbol{k}_{0j'}$ &  $\boldsymbol{k}_{0j'}$ & $\boldsymbol{p}'$ \\
        \hline
    \end{tabular}
    \label{tab:interactionmomentumconfig}
\end{table}

From now on, we specialize to a square 2D lattice and assume that $t^\uparrow = t^\downarrow \equiv t$, $\mu^\uparrow \equiv \mu + \Omega$, $\mu^\downarrow \equiv \mu - \Omega$, $U^{\uparrow\uparrow} = U^{\downarrow\downarrow} \equiv U$ and $U^{\uparrow\downarrow} = U^{\downarrow\uparrow} \equiv \alpha U$. We let the chemical potential $\mu$ control the total number of particles $N$, while the external Zeeman field $\Omega$ controls the spin imbalance, i.e. the values of $N^\alpha$, where $N^\alpha$ is the total number of particles with spin $\alpha$. With the lattice constant $a$ set to $1$, we have $\epsilon_{\boldsymbol{k}} = -2t(\cos k_x + \cos k_y )$ and $s_{\boldsymbol{k}} = -2\lambda_R ( \sin k_y + i \sin k_x )$.

To gain some insight into the SOC BEC, we first discuss the single particle problem, i.e. no interactions. The single particle excitation spectrum is given by the eigenvalues of $\eta_{\boldsymbol{k}}$ which are $\lambda_{\boldsymbol{k}}^\pm =  \epsilon_{\boldsymbol{k}} -\mu \pm \sqrt{\Omega^2+\abs{s_{\boldsymbol{k}}}^2}$. We will refer to these as the upper and lower helicity bands. Let $\Omega_c \equiv 2\lambda_R^2/t$. For $\Omega>\Omega_c$, $\lambda_{\boldsymbol{k}}^-$ has only one minimum at $(0,0)$. For $\Omega< \Omega_c$ it has four minima at $\boldsymbol{k}_{01} = (k_0, k_0), \boldsymbol{k}_{02} = (-k_0, k_0), \boldsymbol{k}_{03} = (-k_0, -k_0)$ and $\boldsymbol{k}_{04} = (k_0, -k_0)$ with $k_0 = k_{0m}$,
\begin{equation}
\label{eq:k0m}
    k_{0m}  \equiv \arcsin\sqrt{(1-(\Omega/\Omega_c)^2)/(1+2(t/\lambda_R)^2)}.
\end{equation}
The same was found in \cite{Toniolo} where both $\Omega > \Omega_c$ and $\Omega< \Omega_c$ were considered for $\alpha < 1$. In this paper we will include $\alpha \geq 1$ and focus on $\Omega < \Omega_c$. 

%In a continuum, Rashba spin-orbit coupling leads to a degenerate ring of minima in the single particle excitation spectrum. With a square lattice, one instead finds four distinct minima as shown above. 

We need to diagonalize the Hamiltonian \eqref{eq:MFTH} in order to obtain the quasiparticle excitation spectrum, and we will consider two methods of obtaining it. One way is to employ the method used in \cite{Toniolo} which involves projecting down on the lowest helicity band. The argument for the validity of the helicity projection is that we are considering a BEC at zero temperature, and so, before introducing interactions, almost no helicity quasiparticles should occupy the upper helicity band. The other method will be to treat the system in the original spin basis, which is equivalent to keeping both helicity bands.

Using the spin basis and following the Bogoliubov approach \cite{bogoliubov1947theory, Bogoliubovvalid}, we insert 
\begin{equation}
\label{eq:BogApp}
    A_{\boldsymbol{k}_{0i}}^\alpha \to \sqrt{N_{0i}^\alpha}e^{-i\theta_{i}^\alpha},
\end{equation}
where $N_{0i}^\alpha = \langle A_{\boldsymbol{k}_{0i}}^{\alpha\dagger} A_{\boldsymbol{k}_{0i}}^\alpha \rangle \gg 1$ is the number of condensate particles with momentum $\boldsymbol{k}_{0i}$ and spin $\alpha$. The angle $\theta_{i}^\alpha$ is a variational parameter that can be determined by minimizing the free energy \cite{bruusflensberg}. It was found that these angles are important in the phases under consideration in this paper. 

We define the helicity operators $C_{\boldsymbol{k}}^+$ and $C_{\boldsymbol{k}}^-$, which annihilate bosons in the upper and lower helicity bands. These are connected to the spin operators $A_{\boldsymbol{k}}^\alpha$ through a unitary matrix containing the eigenvectors of $\eta_{\boldsymbol{k}}$. The eigenvector for the lowest helicity band contains the transformation coefficients 
\begin{equation*}
     u_{\boldsymbol{k}} = \sqrt{\left(1+\Omega/\sqrt{\Omega^2+4\lambda_R^2(\sin^2 k_x + \sin^2 k_y)}\right)/2}
\end{equation*}
and
$v_{\boldsymbol{k}} = -e^{i\gamma_{\boldsymbol{k}}}\sqrt{1-u_{\boldsymbol{k}}^2} $ with $e^{-i\gamma_{\boldsymbol{k}}} \equiv s_{\boldsymbol{k}}/|s_{\boldsymbol{k}}|$.
The helicity projection involves setting $C_{\boldsymbol{k}} \equiv C_{\boldsymbol{k}}^-$ and $C_{\boldsymbol{k}}^+ \approx 0$. Then we have $A_{\boldsymbol{k}}^\uparrow = u_{\boldsymbol{k}} C_{\boldsymbol{k}}$ and $ A_{\boldsymbol{k}}^\downarrow = v_{\boldsymbol{k}} C_{\boldsymbol{k}}$. We transform the Hamiltonian before we use \eqref{eq:BogApp}, and instead insert $C_{\boldsymbol{k}_{0i}} \to \sqrt{N_{0i}}e^{-i\theta_{i}}$, where $N_{0i}$ is the total number of condensate particles with momentum $\boldsymbol{k}_{0i}$. In the helicity projection, we found that the free energy is independent of the angles $\theta_i$, and they are therefore set to zero for brevity. 

\subsection{Phases}
Without interactions, most of the helicity quasiparticles should occupy the minima of $\lambda_{\boldsymbol{k}}^-$. It is expected that introducing weak interactions will designate certain momenta as the ground state \cite{SOCOLRev}, and that a Bogoliubov effect appears such that the condensate momenta become phonon minima of the excitation spectrum, similar to the treatment of the weakly interacting Bose gas \cite{PethickSmith, Pitaevskii, abrikosov}. 

\begin{figure}
    \centering
    \includegraphics[width=\linewidth]{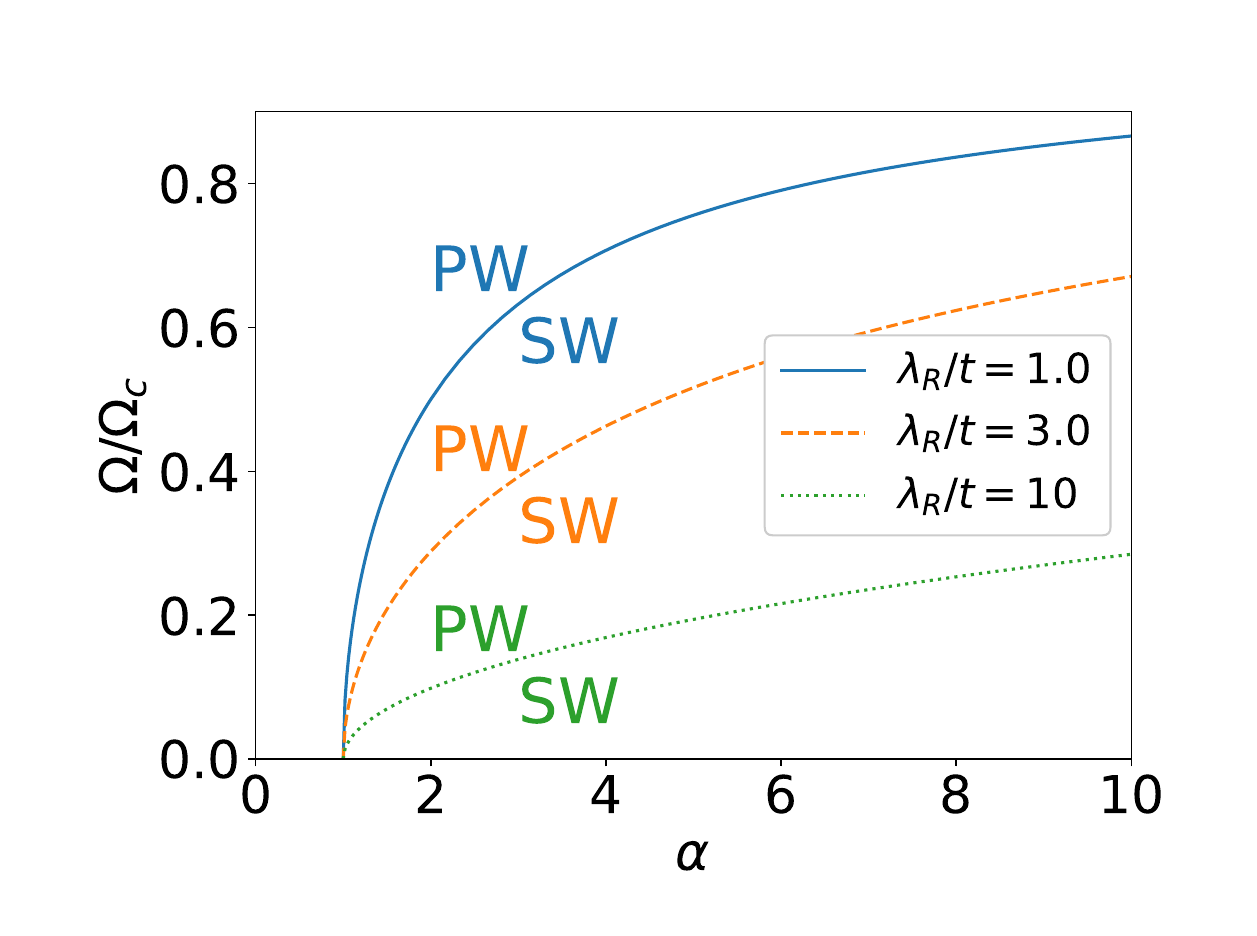} 
    \caption{The PWSW transition line for $U/t = 0.1$, $n = 1$ and $\lambda_R/t = 1.0, 3.0$ and $10$.}
    \label{fig:PWSWtrans}
\end{figure}

As is often done \cite{Toniolo, SOCOLRev}, we will use the operator independent part, $H_0$, of the Hamiltonian to determine the possible phases. We are thus assuming that the free energy $F \approx H_0$, and the phase with the lowest free energy at a certain set of parameters will be the preferred phase. With a nonzero SOC and $\Omega<\Omega_c$ the two most interesting phases are the plane and stripe wave phases, named according to the wave patterns they produce in real space, and characterized by
\begin{itemize}
    \item Plane Wave (PW) Phase: The PW phase involves a single nonzero condensate momentum, chosen to be $\boldsymbol{k}_{01}$ without loss of generality. 
    \item Stripe Wave (SW) Phase: The SW phase involves condensation at two oppositely directed, nonzero momenta chosen as $\boldsymbol{k}_{01}$ and $\boldsymbol{k}_{03} = -\boldsymbol{k}_{01}$.
\end{itemize}
When $\Omega = 0$ the PWSW transition occurs at $\alpha = 1$. For $0<\Omega<\Omega_c$, the transition occurs at \cite{Toniolo}
\begin{equation}
\label{eq:PWSWtransition}
    \Omega/\Omega_c = \sqrt{(\alpha-1)/(\alpha+1+(\lambda_R/t)^2)}.
\end{equation}
This analytic expression was found using the operator independent part of the Hamiltonian after projecting onto the lowest helicity band, and is found to be an adequate approximation. The SW phase is preferred for $\Omega$ less than the value given above and its excitation spectrum was not treated in \cite{Toniolo}. A plot of this transition line is shown in figure \ref{fig:PWSWtrans}. 
See figure 1 in \cite{Toniolo} for an $\Omega-\alpha$ phase diagram based on $H_0$.
% See figure \ref{fig:PWSWPD} for an $\Omega-\alpha$ phase diagram, focusing on $\Omega < \Omega_c$.

The obtained PWSW transition at $\alpha=1$ when $\Omega = 0$ was also found in \cite{wangPWSWTransition}. As further elaborated in \cite{zhaiSWRashbaRev, SOCOLRev} the wave function in the PW phase gives a uniform density of both spin components, while in the SW phase both components have a periodic, striped density variation with opposite phase. Since this minimizes the overlap of the two components, the SW phase is preferred when the intercomponent interactions are stronger than the intracomponent interactions. Upon introducing a Zeeman field the system obtains a spin imbalance. Hence, minimizing the overlap of the two components becomes less effective, and a higher value of $\alpha$ is required before the SW phase is energetically favorable.

\subsection{Generalized Diagonalization Method}
Since the system is bosonic and the Hamiltonian contains terms that individually do not conserve the particle number, diagonalizing the Hamiltonian \eqref{eq:MFTH} must be done by a generalization of the Bogoliubov-Valatin (BV) transformation introduced in \cite{Tsallis} with further details in \cite{Xiao}. A unitary transformation is in general not sufficient for a bosonic system as there is no guarantee the quasiparticles will be bosonic. To circumvent this, the BV transformation introduces a matrix $J = \begin{pmatrix} I & 0 \\ 0&-I \end{pmatrix}$,
where $I$ is the identity matrix. If $M$ is the matrix in a Hamiltonian which is quadratic in bosonic operators, then the BV transformation involves diagonalizing $MJ$. Complex eigenvalues of $MJ$ are interpreted as dynamic instabilities of the system \cite{PethickSmith, Pitaevskii}.

%---------------------------------PW PHASE----------------------------------------------------------------------------
\section{Plane Wave Phase} \label{sec:PW}

The PW phase is treated by the helicity projection in \cite{Toniolo}. The quadratic part of the Hamiltonian can be written
\begin{equation}
    H_2 = \frac{1}{4}\sum_{\boldsymbol{k}\neq \boldsymbol{k}_{01}} \boldsymbol{C}_{\boldsymbol{k}}^\dagger N_{\boldsymbol{k}} \boldsymbol{C}_{\boldsymbol{k}},
\end{equation}
where, with $\boldsymbol{p} = 2\boldsymbol{k}_{01}-\boldsymbol{k}$,
\begin{equation}
    \boldsymbol{C}_{\boldsymbol{k}} = (C_{\boldsymbol{k}}, C_{\boldsymbol{p}}, C_{\boldsymbol{k}}^\dagger, C_{\boldsymbol{p}}^\dagger)^T
\end{equation}
and
\begin{gather}
    N_{\boldsymbol{k}} = 
    \begin{pmatrix}
    N_{11}(\boldsymbol{k}) & 0 & 0 & N_{32}^* (\boldsymbol{k}) \\
    0 &  N_{11}(\boldsymbol{p}) & N_{32}^* (\boldsymbol{k}) & 0 \\
    0 & N_{32} (\boldsymbol{k}) & N_{11}(\boldsymbol{k})  & 0 \\
    N_{32} (\boldsymbol{k}) & 0 & 0 & N_{11}(\boldsymbol{p})  \\
    \end{pmatrix}.
\end{gather}
The matrix elements are
\begin{equation}
    \begin{aligned}
    N_{11}(\boldsymbol{k}) &= \lambda_{\boldsymbol{k}}^- - \lambda_{\boldsymbol{k}_{01}}^- +Un\big[ 2u_{\boldsymbol{k}}^2u_{\boldsymbol{k}_{01}}^2 + 2\abs{v_{\boldsymbol{k}}}^2\abs{v_{\boldsymbol{k}_{01}}}^2   \\
    & -u_{\boldsymbol{k}_{01}}^4 - \abs{v_{\boldsymbol{k}_{01}}}^4 \big] + U\alpha n\big[ u_{\boldsymbol{k}}^2\abs{v_{\boldsymbol{k}_{01}}}^2 \\
    &+u_{\boldsymbol{k}_{01}}^2\big(\abs{v_{\boldsymbol{k}}}^2-2\abs{v_{\boldsymbol{k}_{01}}}^2\big) \\
    & + 2u_{\boldsymbol{k}}u_{\boldsymbol{k}_{01}}\Re(v_{\boldsymbol{k}}v_{\boldsymbol{k}_{01}}^* )\big], \\
    N_{32}(\boldsymbol{k}) &= Un\left(u_{\boldsymbol{k}_{01}}^2u_{\boldsymbol{k}}u_{\boldsymbol{p}} + v_{\boldsymbol{k}_{01}}^{*2} v_{\boldsymbol{k}}v_{\boldsymbol{p}}\right) \\
    &+U\alpha n u_{\boldsymbol{k}_{01}}v_{\boldsymbol{k}_{01}}^* \left(u_{\boldsymbol{k}}v_{\boldsymbol{p}} + u_{\boldsymbol{p}}v_{\boldsymbol{k}}\right),
    \end{aligned}
\end{equation}
where $n = N/N_s$. The eigenvalues of $N_{\boldsymbol{k}}J$ are
\begin{align}
    \begin{split}
    \label{eq:PWOMH}
    E_H(\boldsymbol{k}) &= \frac{1}{2}\bigg(N_{11}(\boldsymbol{k}) - N_{11}(\boldsymbol{p})\\
    &+ \sqrt{\big(N_{11}(\boldsymbol{k}) + N_{11}(\boldsymbol{p}) \big)^2 - 4\abs{N_{32} (\boldsymbol{k})}^2}  \bigg),
    \end{split}
\end{align}
and its inverse about $\boldsymbol{k}_{01}$. This agrees with the result obtained in \cite{Toniolo}. Using this inversion symmetry, it is possible to write the diagonalized Hamiltonian as \cite{KM}
\begin{equation}
    H_2 = \sum_{\boldsymbol{k}\neq \boldsymbol{k}_{01}} E_H(\boldsymbol{k}) \left(B_{\boldsymbol{k}}^{\dagger}B_{\boldsymbol{k}} +\frac12 \right) .
\end{equation}

\begin{figure}
    \centering
    \includegraphics[width=\linewidth]{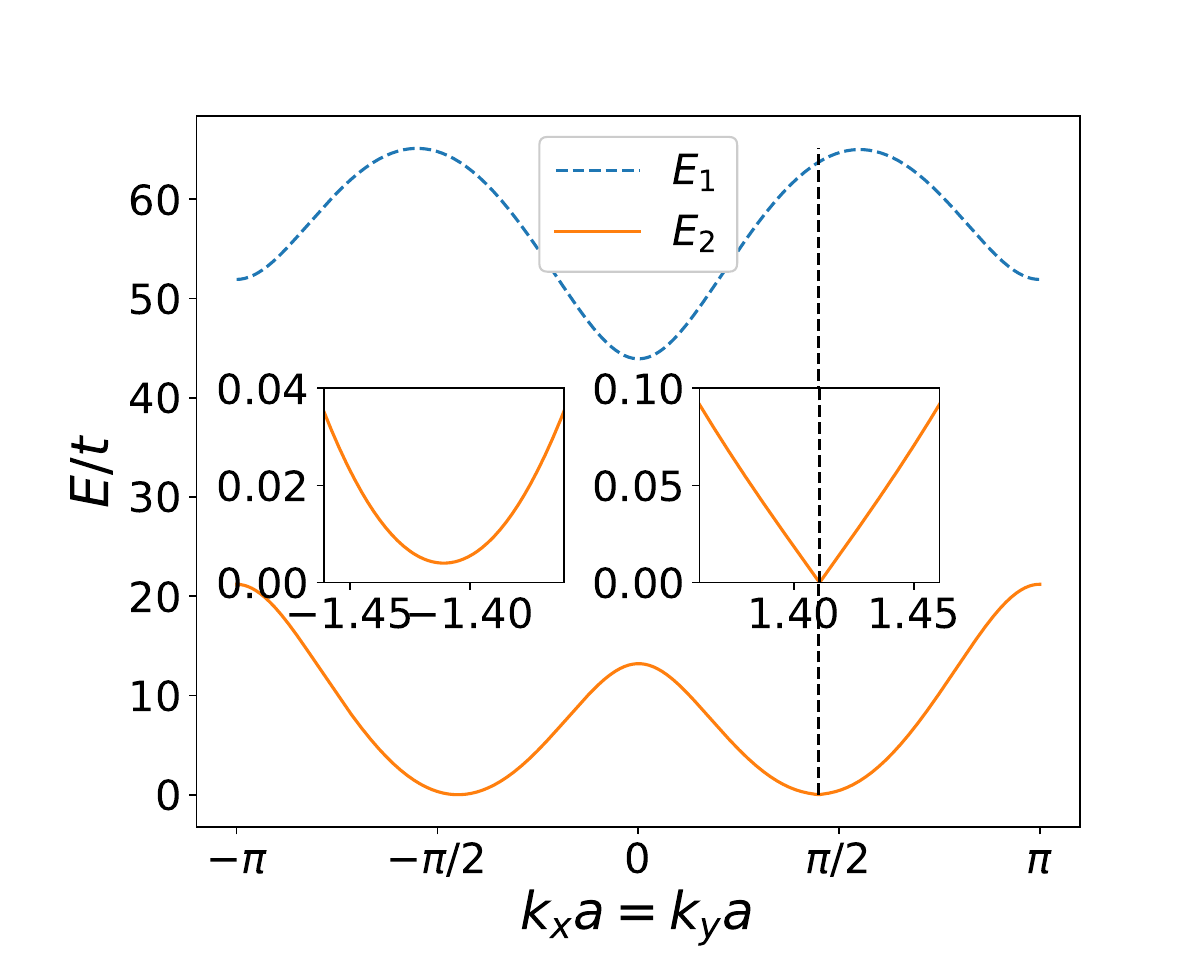}
    \caption{PW phase excitation spectrum in the direction $k_x = k_y$, obtained numerically in the spin basis. The dashed vertical line shows the position of $\boldsymbol{k} = \boldsymbol{k}_{01}$. The insets show the phonon minimum at the condensate momentum, and the gapped roton minimum close to $-\boldsymbol{k}_{01}$. The parameters were set to $U/t = 0.1$, $n = 1$, $\lambda_R/t = 10$, $\alpha = 1.5$ and $\Omega/t \approx 15.37$. This value of $\Omega$ corresponds to $1.1$ times the PWSW transition line \eqref{eq:PWSWtransition}.}
    \label{fig:PWband}
\end{figure}

As discussed in \cite{Toniolo} this energy band has a phonon minimum at the condensate momentum, $\boldsymbol{k}_{01}$, and gapped roton minima close to the other minima of the single particle excitation spectrum. This is illustrated in the insets of figure \ref{fig:PWband}. When approaching the PWSW transition line \eqref{eq:PWSWtransition} from above, the roton minimum close to $\boldsymbol{k}_{03}$ goes to zero, and eventually becomes negative, indicating an energetic instability \cite{PethickSmith}. 

Treating the PW phase in the spin basis requires a numerical solution for the eigenvalues of an $8 \cross 8$ matrix. The method follows the same course as the SW phase, to be presented later, and is therefore omitted here. The matrix elements are shown in appendix \ref{sec:PWmatelem}. The lowest band, $E_2(\boldsymbol{k})$, is almost equal to the eigenvalue $E_H(\boldsymbol{k})$ in the helicity projection at all $\boldsymbol{k}$, while the upper band, $E_1(\boldsymbol{k})$, is similar to the upper helicity band $\lambda_{\boldsymbol{k}}^+ (\boldsymbol{k})$. Both bands are shown in figure \ref{fig:PWband}. Using $E_H(\boldsymbol{k})$ we can find an analytic expression for the anisotropic sound velocity of the excitations close to $\boldsymbol{k}_{01}$ \cite{Toniolo}. The numerical sound velocity from the spin basis corresponds to this analytic result, even without any Zeeman field and at weak SOC. Hence, the helicity projection provides a good approximation for the PW phase at all parameters of interest, even though it is expected to be a better approximation at strong SOC and with a Zeeman field $\Omega > \max\{U, \alpha U\}$ \cite{Toniolo}. The latter requirement is intended to reduce interband scatterings between the helicity bands. Apparently, the interband scatterings are not relevant for the speed of sound of the phonon excitations in the PW phase.

%----------------------------------SW PHASE-----------------------------------------------
\section{Stripe Wave Phase} \label{sec:SW}
\subsection{Spin basis}
Since there are no terms in the Hamiltonian that would introduce a momentum imbalance, we assume that $N_{\boldsymbol{k}_{01}}^\uparrow = N_{\boldsymbol{k}_{03}}^\uparrow = N_0^\uparrow/2$ and $N_{\boldsymbol{k}_{01}}^\downarrow =  N_{\boldsymbol{k}_{03}}^\downarrow = N_0^\downarrow/2$, where $N_0^\alpha$ is the total number of condensate particles with spin $\alpha$. Using \eqref{eq:H0} and \eqref{eq:BogApp} we find an initial expression for $H_0$. Then we use $N^\alpha = N_0^\alpha + \sum_{\boldsymbol{k}}^{'} A_{\boldsymbol{k}}^{\alpha\dagger} A_{\boldsymbol{k}}^{\alpha}$ to replace $N_0^\alpha$ by $N^\alpha$. In this replacement we neglect terms that are more than quadratic in excitation operators. The new operator independent part is named $H_0$, while the terms that are quadratic in excitation operators are moved to $H_2$. We may replace $N_0^\alpha$ by $N^\alpha$ directly in $H_1$ and $H_2$ to the same order of approximation. The treatment of the linear terms in $H_1$ is only relevant for a calculation of the free energy, which is left for appendix \ref{sec:SWF}. 

We set $N^\uparrow = Nx$ and $N^\downarrow = N(1-x)$, with $1/2 \leq x \leq 1$ when $\Omega \geq 0$. We view $x$ as a variational parameter that can be determined by minimizing the free energy. The expression for $H_0$ is
\begin{align}
    \begin{split}
        H_0 &= (\epsilon_{\boldsymbol{k}_{01}}-\mu)N + \Omega N (1-2x)\\
        &+N\sqrt{x(1-x)}\abs{s_{\boldsymbol{k}_{01}}}\sum_{i=1,3}\cos(\gamma_{\boldsymbol{k}_{0i}}+\Delta\theta_i) \\ 
        & +\frac{UN^2}{4N_s}\Big(3x^2 + 3(1-x)^2 \\
        &\mbox{\qquad} +2\alpha x(1-x)\big(2+\cos(\Delta\theta_1-\Delta\theta_3)\big)\Big),
    \end{split}
\end{align}
where $\Delta\theta_i \equiv \theta_i^\downarrow - \theta_i^\uparrow$.
% Finding the minimum of $H_0$ w.r.t $x$ is done numerically, since the eigenvalues of $M_{\boldsymbol{k}}J$ can only be found numerically anyway.

We write $H_2$ as
\begin{equation}
    H_2 = \frac{1}{4}\left.\sum_{\boldsymbol{k}}\right.^{'}\boldsymbol{A}_{\boldsymbol{k}}^{\dagger} M_{\boldsymbol{k}}\boldsymbol{A}_{\boldsymbol{k}}.
\end{equation}
Introducing $\boldsymbol{p}_\pm = \boldsymbol{k} \pm 2\boldsymbol{k}_{01}$ and $\boldsymbol{q}_\pm = -\boldsymbol{k} \pm 2\boldsymbol{k}_{01}$, the operator vector is defined by
\begin{align}
    \begin{split}
    \label{eq:SWbra}
        \boldsymbol{A}_{\boldsymbol{k}}^{\dagger} = (&A_{\boldsymbol{k}}^{\uparrow\dagger}, A_{-\boldsymbol{k}}^{\uparrow\dagger}, A_{\boldsymbol{p}_+}^{\uparrow\dagger}, A_{\boldsymbol{q}_+}^{\uparrow\dagger}, A_{\boldsymbol{p}_-}^{\uparrow\dagger}, A_{\boldsymbol{q}_-}^{\uparrow\dagger}, \\
        &A_{\boldsymbol{k}}^{\downarrow\dagger}, A_{-\boldsymbol{k}}^{\downarrow\dagger}, A_{\boldsymbol{p}_+}^{\downarrow\dagger}, A_{\boldsymbol{q}_+}^{\downarrow\dagger}, A_{\boldsymbol{p}_-}^{\downarrow\dagger}, A_{\boldsymbol{q}_-}^{\downarrow\dagger}, \\
        &A_{\boldsymbol{k}}^{\uparrow}, A_{-\boldsymbol{k}}^{\uparrow}, A_{\boldsymbol{p}_+}^{\uparrow}, A_{\boldsymbol{q}_+}^{\uparrow}, A_{\boldsymbol{p}_-}^{\uparrow}, A_{\boldsymbol{q}_-}^{\uparrow}, \\
        &A_{\boldsymbol{k}}^{\downarrow}, A_{-\boldsymbol{k}}^{\downarrow}, A_{\boldsymbol{p}_+}^{\downarrow}, A_{\boldsymbol{q}_+}^{\downarrow}, A_{\boldsymbol{p}_-}^{\downarrow}, A_{\boldsymbol{q}_-}^{\downarrow}),
    \end{split}
\end{align}
and $M_{\boldsymbol{k}}$ is a $24 \cross 24$ matrix on the form
\begin{gather}
\label{eq:SWMat}
    M_{\boldsymbol{k}} = 
    \begin{pmatrix}
    M_1 & M_2 \\
    M_2^* & M_1^* \\
    \end{pmatrix}.
\end{gather}
The matrix elements are presented in appendix \ref{sec:SWmatelem}. They are obtainable from the expression \eqref{eq:H2}, by using commutators and making $-\boldsymbol{k}$ terms explicit, a procedure that produces some additional operator independent terms in the Hamiltonian, relevant for a calculation of the free energy. More details are found in appendix \ref{sec:SWF}.

\begin{figure}
    \centering
    \includegraphics[width=\linewidth]{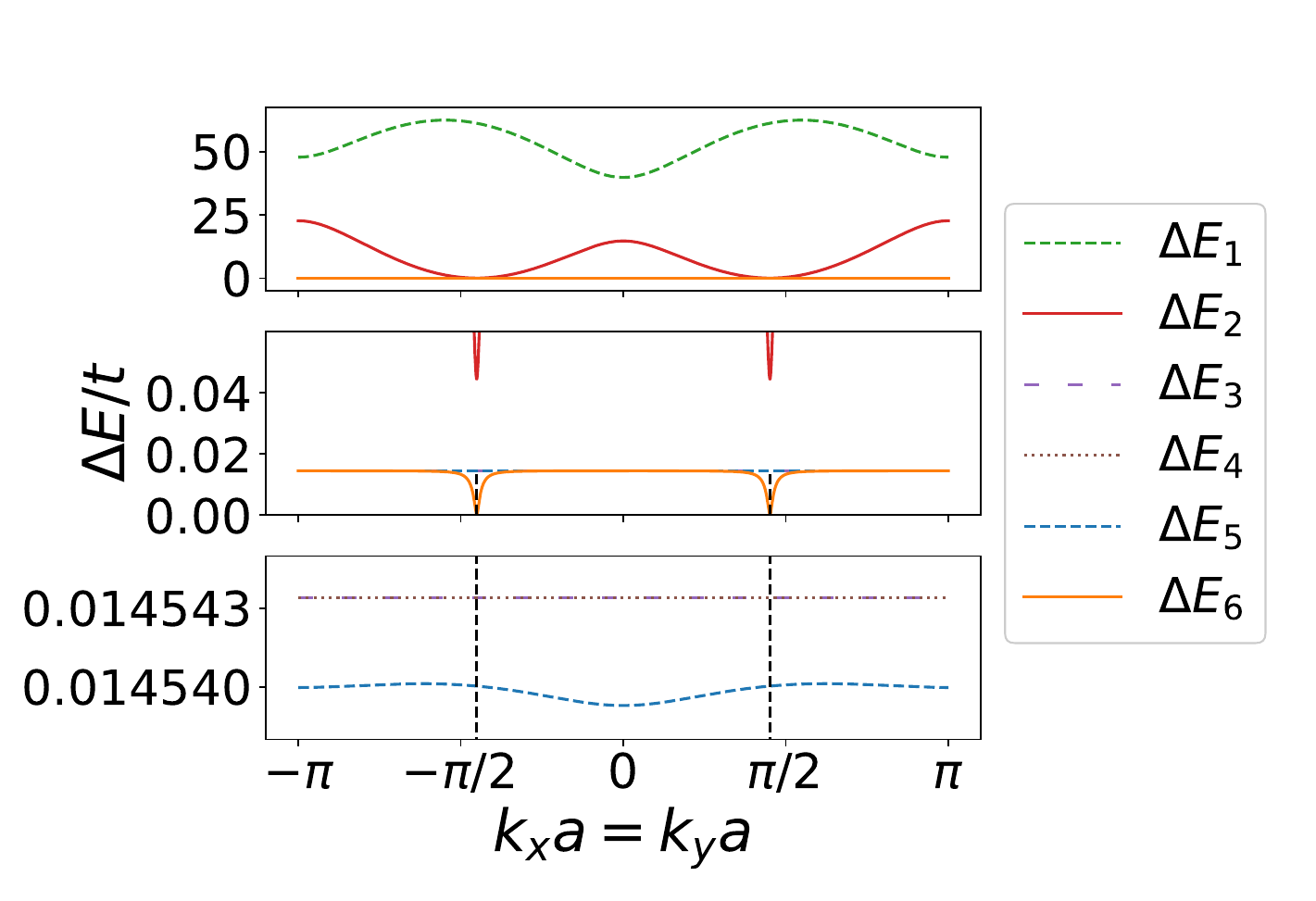}
    \caption{The energy bands in the SW phase in the direction $k_x = k_y$. The dashed vertical lines show the positions of $\boldsymbol{k} = \pm\boldsymbol{k}_{01}$. The parameters were set to $U/t = 0.1$, $n = 1$, $\lambda_R/t = 10$, $\alpha = 1.5$ and $\Omega/t \approx 12.57$. This value of $\Omega$ corresponds to $0.9$ times the PWSW transition line \eqref{eq:PWSWtransition}.}
    \label{fig:SWband}
\end{figure}

\begin{figure}
    \centering
    \begin{minipage}{.24\textwidth}
      \includegraphics[width=\linewidth]{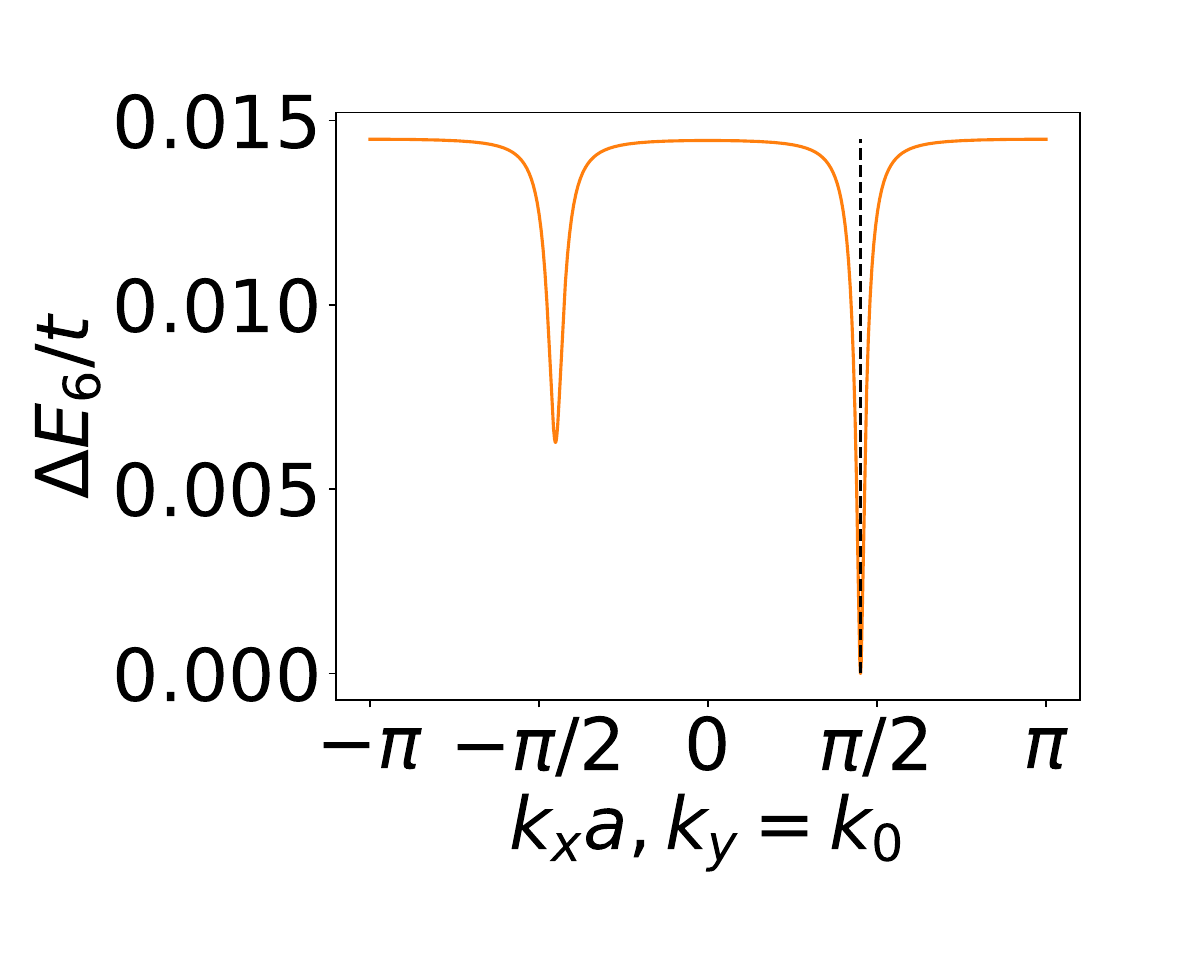}
      \label{fig:SWeigen}
    \end{minipage}%
    \begin{minipage}{.24\textwidth}
      \includegraphics[width=\linewidth]{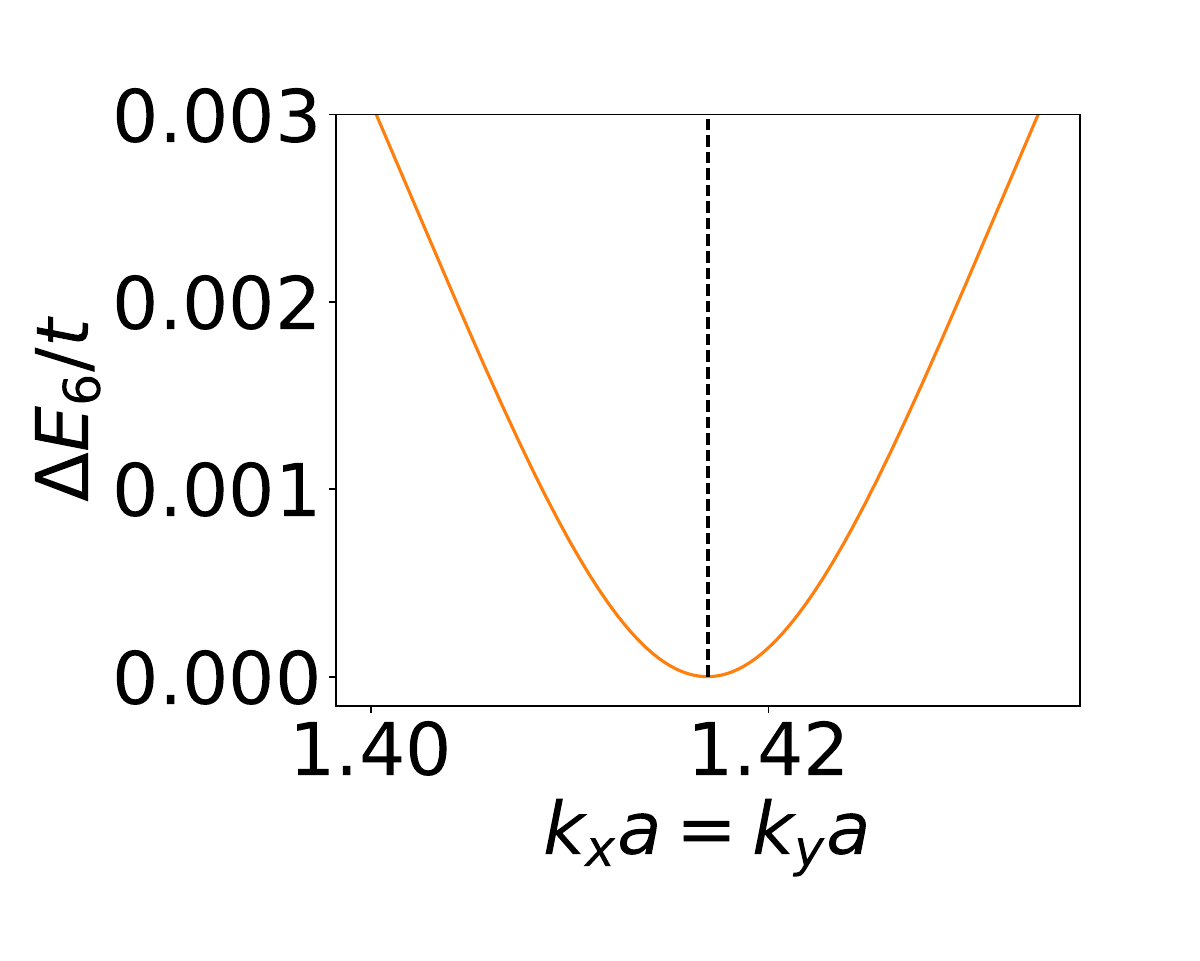}
      \label{fig:SWmin}
    \end{minipage}%
    \caption{(Left) The lowest energy band in the SW phase shown along $k_x$ for $k_y = k_0$. (Right) The quadratic minimum close to $\boldsymbol{k}_{01}$ for the lowest energy band in the direction $k_x = k_y$. The dashed vertical lines show the position of $\boldsymbol{k} = \boldsymbol{k}_{01}$. The parameters are the same as in figure \ref{fig:SWband}.  \label{fig:SWexcitation}}
\end{figure}

The 24 eigenvalues of $M_{\boldsymbol{k}}J$ are equally distributed around 0 \cite{Tsallis, Xiao}. Eight eigenvalues are within numerical accuracy 0, while the remaining eigenvalues are doubly degenerate, upon inserting the values of the variational parameters which minimize the free energy. The two lowest positive, doubly degenerate eigenvalues have anomalous modes \cite{PethickSmith}, and therefore enter the diagonalized Hamiltonian with a negative sign \cite{Xiao}. By moving the chemical potential controlling the quasiparticles to just below the (negative) minimum of the excitation spectrum, $-E_0$, Bose-Einstein statistics ensure that the majority of the quasiparticles will occupy the minima of the lowest band. Since we prefer to have only positive energy bands, we move the zero of energy by $E_0$. The final diagonalized Hamiltonian reads \cite{KM}
\begin{equation}
     H_2 = -E_0 N_q +  \left.\sum_{\boldsymbol{k}}\right.^{'}\sum_{\sigma=1}^{6} \Delta E_\sigma(\boldsymbol{k})\left(B_{\boldsymbol{k},\sigma}^{\dagger}B_{\boldsymbol{k},\sigma}+\frac12\right).
\end{equation}
where the quantity 
$N_q \equiv \sum_{\boldsymbol{k}}^{'}\sum_{\sigma=1}^{6}(B_{\boldsymbol{k},\sigma}^{\dagger}B_{\boldsymbol{k},\sigma}+1/2)$
was defined to simplify the expression, and $\Delta$ is used to indicate that the energies have been shifted by $E_0$. The energies are ordered such that $\Delta E_i(\boldsymbol{k}) \geq \Delta E_j(\boldsymbol{k})$ if $j \geq i$ and are shown in figure \ref{fig:SWband}. Since there are only two degrees of freedom originally, spin up and spin down, the four highest excitation energies will be considered unoccupied. The lowest energy band, $\Delta E_6 (\boldsymbol{k})$ is the most interesting band in the context of BEC and is shown in figure \ref{fig:SWexcitation}. It has its global minima at the condensate momenta, and gapped roton minima at the unoccupied minima of the single particle spectrum. Note the highly unusual feature that, unlike the typical results when introducing interactions, the minima at the condensate momenta show a non-linear behavior. 

These quadratic minima indicate that the excitations in the SW phase have zero sound velocity, and separate our results from those of continuum SW phase excitation spectra studied in \cite{li2013superstripes, Swexcitation3Dcont}, wherein nonzero sound velocities are found. The SW phase excitation spectrum, found numerically in the spin basis, is the main result of this paper. In the next subsection we consider the helicity projection, and find that it is a poor approximation in the SW phase. The treatment is however useful, since it provides a way to explain the quadratic behavior found in the SW phase. 

\subsection{Helicity Projection}
Obtaining the quasiparticle excitation spectrum utilizing the helicity projection, follows the same course as in the spin basis, the difference being the projection onto the lowest helicity band. This reduces the number of components of the basis to $12$. The matrix is presented in appendix \ref{sec:SWmatelemhel}.

The excitation energies are, at first glance, similar to the energies $\Delta E_{2i}, i=1,2,3$ in the spin basis. We denote them $\Delta E_\sigma^H, \sigma = 1,2,3$. The lowest band $\Delta E_3^H(\boldsymbol{k})$ however, has some properties that separates it from the spin basis result $\Delta E_{6}(\boldsymbol{k})$. At zero Zeeman field and $\lambda_R/t < \sqrt{6}$, the minima at the condensate momenta show a linear behavior, contrary to the result in the spin basis, but more in accord with the intuition one would have based on a standard single-momentum condensate. However, for $\lambda_R/t > \sqrt{6}$ and $\Omega = 0$ or any $\lambda_R$ with nonzero $\Omega$, a quadratic behavior is found.

The value $\lambda_R/t = \sqrt{6}$ corresponds to $k_0 = \pi/3$. This is the point where $\boldsymbol{k}+2\boldsymbol{k}_{01}$ at $\boldsymbol{k} = \boldsymbol{k}_{01}$ goes beyond the 1BZ. The term $s_{\boldsymbol{k}}/|s_{\boldsymbol{k}}|$ involved in the transformation to the helicity basis has discontinuities when $\boldsymbol{k}$ crosses the boundary of the 1BZ. For $\lambda_R/t < \sqrt{6}$ and $\Omega = 0$ certain matrix elements are zero around $\boldsymbol{k}_{01}$ or $-\boldsymbol{k}_{01}$, while they become nonzero when $\lambda_R/t > \sqrt{6}$. This appears to be the root cause for why the linear behavior of the excitation spectrum is replaced by quadratic behavior. 

\begin{figure}
    \centering
    \includegraphics[width=\linewidth]{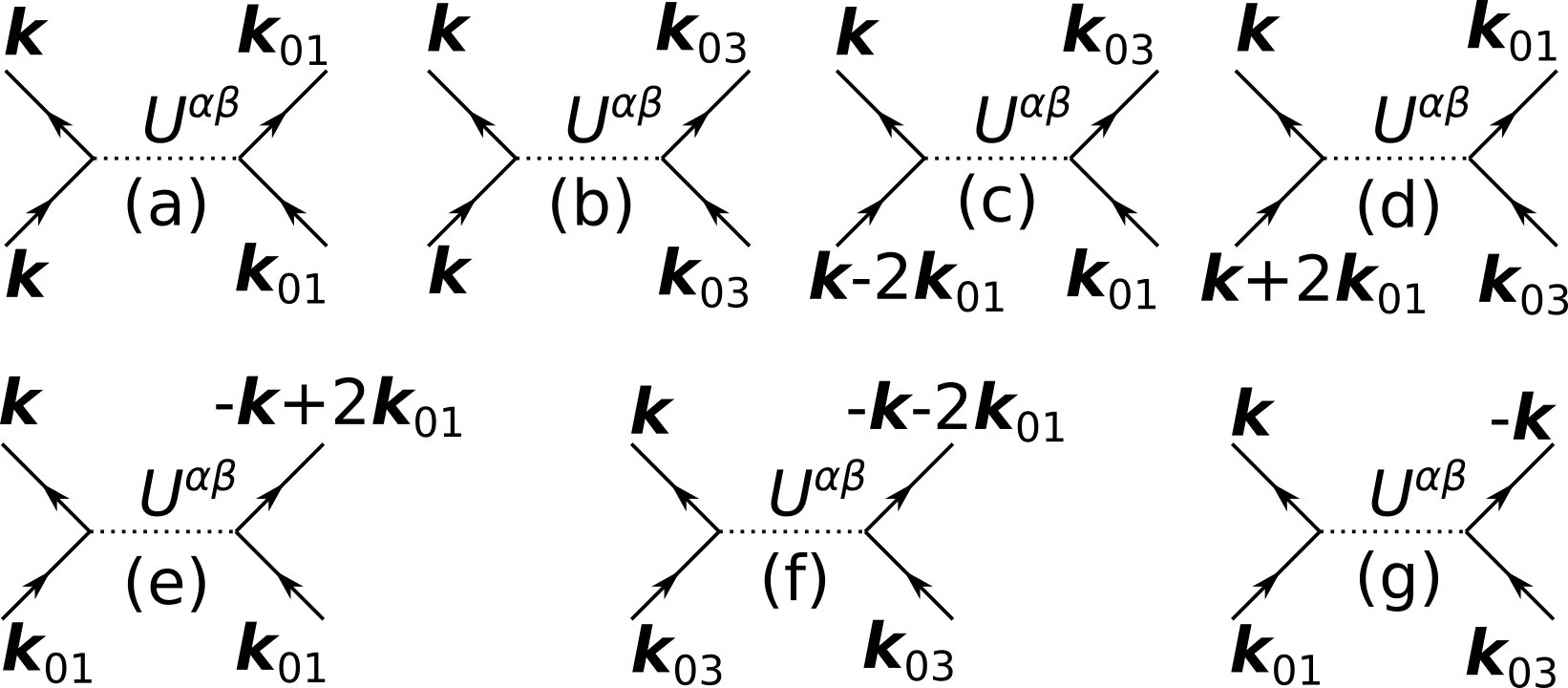}
    \caption{The figure represents all the scattering processes relevant for the quadratic part of the Hamiltonian in the SW phase. Time progresses upwards, and the reverse processes are also relevant. Interchanging either the incoming or the outgoing momenta reveals other relevant scatterings. The dotted line illustrates the on-site interaction with strength $U^{\alpha\beta}$ when the two particle lines have spin $\alpha$ and $\beta$ respectively.}
    \label{fig:SWscatterings}
\end{figure}

The absolute squares of these matrix elements represent the transition rates of the scatterings (c), (d), (e) and (f) in figure \ref{fig:SWscatterings}. There is no reason why the transition rates of these scatterings should be zero, something which is supported by the fact that they are nonzero in the original spin basis. The conclusion is that the helicity projection fails for weak SOC and zero Zeeman field due to the discontinuities in the transformation to the helicity basis with zero Zeeman field. Furthermore, the helicity projection should be a better approximation at stronger SOC, where it too shows quadratic behavior even without a Zeeman field. However, it is found that when $\lambda_R/t > \sqrt{6}$, the global minima of the excitation spectrum occur at $\pm\boldsymbol{k}_{02}$, instead of at $\pm\boldsymbol{k}_{01}$ as in the spin basis. Since the spin basis is more accurate than the helicity projection, we conclude that the helicity projection fails in the SW phase at almost all parameters, the possible exception being for $\lambda_R/t < \sqrt{6}$ and $\Omega>0$.

The main reason why the discontinuities in the transformation to the helicity basis when $\Omega = 0$ have such a large influence on the SW phase, but apparently no influence on the PW phase, is the presence of two condensate momenta. In the PW phase, the momentum indices of the operators are $\boldsymbol{k}$ and $\boldsymbol{p} = 2\boldsymbol{k}_{01}-\boldsymbol{k}$ only. At the condensate momentum, $\boldsymbol{p} = \boldsymbol{k}_{01}$ and there are no problems with this crossing the boundary of the 1BZ since the condensate momentum is kept in the 1BZ by definition. On the other hand, the presence of two condensate momenta enables more scattering processes such that e.g. $\boldsymbol{p}_+ = \boldsymbol{k} + 2\boldsymbol{k}_{01}$ becomes one of the momentum indices in the operators. Hence, it is possible for the discontinuities of the transformation to the helicity basis to directly influence the excitation spectrum close to the condensate momenta.   

The remaining question is why the excitation spectrum in the SW phase shows quadratic behavior close to its minima, contrary to the usual Bogoliubov result when introducing interactions. The simplest explanation is that it is a consequence of the presence of more than one condensate momentum, a situation which has no counterpart in the standard treatments of such interacting condensates. The presence of two condensate momenta is the reason for the large basis, and the number of nonzero matrix elements. Furthermore, removing a certain set of these matrix elements is required to obtain a linear result. We further discuss how interactions generally lead to linear minima, and how the SW phase breaks with the conventional behavior in appendix \ref{sec:linearquadratic}.

\subsection{Stability}
With $\Omega = 0$ it is found that the SW phase is stable when $\alpha>1$ \cite{KM}. Introducing a Zeeman field, we find that on approaching the PWSW transition line \eqref{eq:PWSWtransition} from below, the excitation spectrum in the SW phase becomes complex, indicating a dynamical instability. Like the energetic instability of the PW phase, this occurs very close to the PWSW transition line. Hence, there is a small area close to this line where neither phase is stable. It may be of interest to study which phase the system will enter in this area. The main candidate is the lattice wave (LW) phase involving all four condensate momenta \cite{master, KM}, which did not enter the phase diagram when neglecting excitations \cite{Toniolo}. This paper will not explore this further.

%x^PW neq x^SW

When $\lambda_R/t = 1.0$ and $\alpha \gtrsim 2$ we find that energetic instabilities develop for $\Omega$ around approximately half the PWSW transition line and beyond $\sim 0.9$ of the PWSW transition line. A greater set of values for $\Omega$ is affected by these instabilities when $\alpha$ is increased. These energetic instabilities are characterized by a distance between the minima of the excitation spectrum and the condensate momenta considered to be so large that these no longer correspond to the same lattice sites. If the minima of the excitation spectrum are not located at the condensate momenta, then the initial assumption that the system condenses at $\pm\boldsymbol{k}_{01}$ is invalid. Once again, this paper will not explore the system in the region where neither the PW nor the SW phase is stable. We  note that upon choosing $\lambda_R/t = 10$, these energetic instabilities disappear inside the region where the SW phase is already dynamically unstable. One can understand this behavior by considering figure \ref{fig:PWSWtrans} showing the PWSW transition line at different SOC strengths. We notice that when the strength of SOC increases, the maximum value of $\Omega/\Omega_c$ found in the SW phase decreases. Hence, a value of $\Omega$ close to the PWSW transition line when $\lambda_R/t = 10$ represents a significantly smaller $\Omega/\Omega_c$ than when $\lambda_R/t = 1.0$.

A calculation of the ground state depletion for a weakly interacting Bose gas can be found in e.g. \cite{Pitaevskii}. The calculation here is completely analogous, except now we must use the numerically constructed BV transformation matrix \cite{Tsallis, Xiao, KM}. At zero temperature and for parameters where the SW phase is stable, we find that $(N-N_0)/N$ is lower than $1 \%$ when $U/t = 0.1$, confirming the validity of the mean-field theory.

\begin{figure*}
    \centering
    \includegraphics[width=\linewidth]{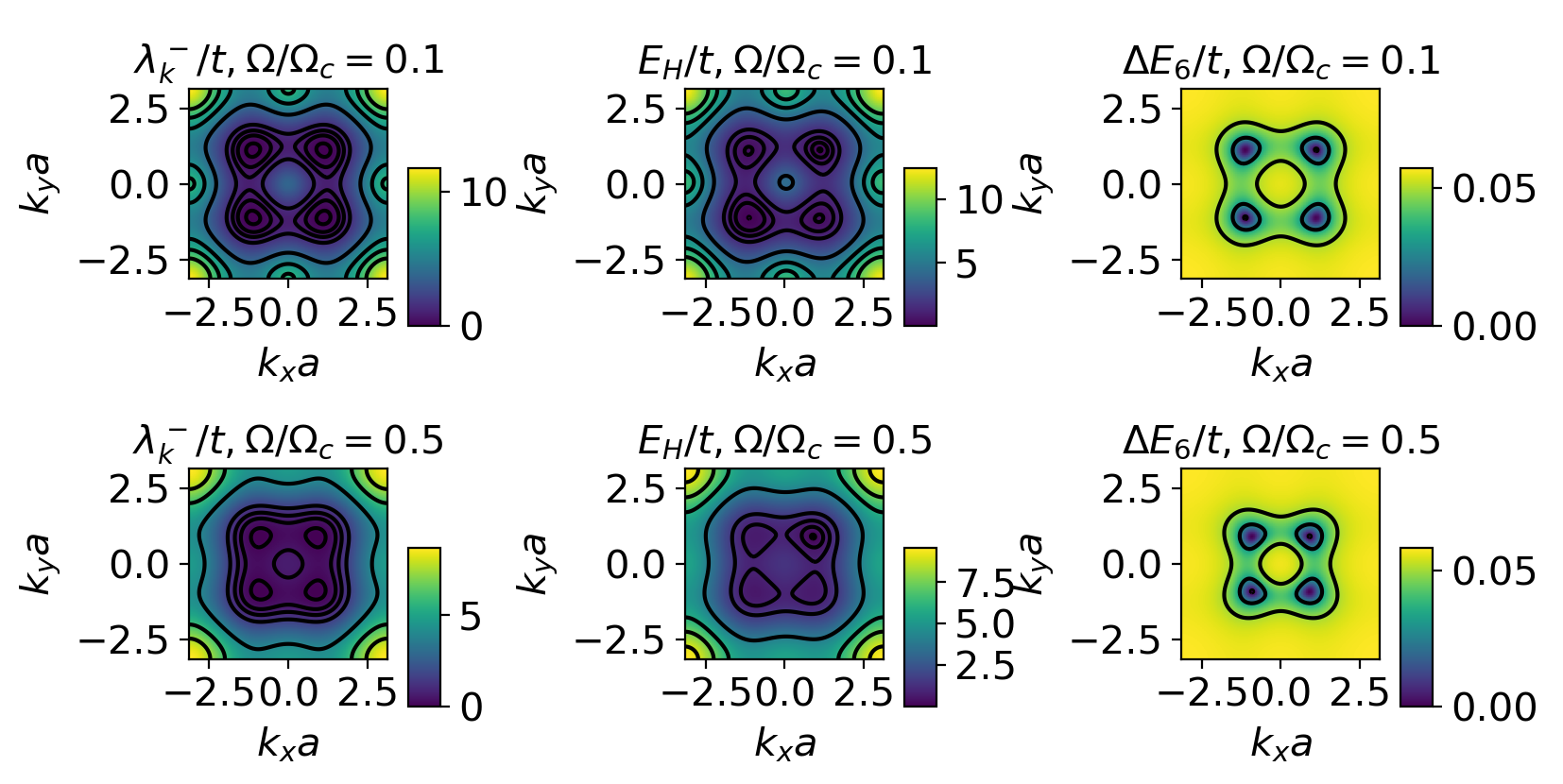}
    \caption{The lowest bands in the 1BZ for the single particle problem, $\lambda_{\boldsymbol{k}}^-$, for the PW phase using the helicity projection, $E_H(\boldsymbol{k})$, and for the SW phase using the spin basis, $\Delta E_6(\boldsymbol{k})$. The black lines are contour lines. Two values of $\Omega/\Omega_c < 1$ are considered, and $\lambda_R/t = 3.0$ in all figures. The lowest helicity band has been shifted so that its minimum is zero. In the PW phase we used $U/t = 1.0$ and $\alpha = 0.9$ to better show its difference from the lowest helicity band. In the SW phase, $U/t = 0.1$ and $\alpha = 10$ were used to ensure stability at both values of $\Omega/\Omega_c$. Note the four-fold symmetry of the helicity bands, reflecting the underlying four-fold symmetry of the optical lattice.}
    \label{fig:lowband}
\end{figure*}

Though we have calculated all bands for the single particle problem and the two phases, the most important bands in the context of BEC are the lowest bands. We summarize our results in figure \ref{fig:lowband} showing the lowest bands in the 1BZ. As we can also see from \eqref{eq:k0m}, the length of the momenta at the minima decrease when $\Omega$ increases. The lowest band in the PW phase is very similar to the lowest helicity band except for the phonon minimum at the condensate momentum and gapped roton minima at the other three momenta where $\lambda_{\boldsymbol{k}}^-$ has its minima. The figure also reveals that the SW phase is special, since the maximum value of its lowest band is much lower than the lowest helicity band. This is connected to the presence of both zero and anomalous modes in the SW phase excitation spectrum. These are caused by the presence of two condensate momenta, enabling a larger set of possible scatterings and so a larger matrix from which the excitation spectrum is obtained. The presence and location of the four local minima is however very similar to the lowest helicity band. There are two quadratic minima at the condensate momenta, and gapped roton minima at the other two momenta where $\lambda_{\boldsymbol{k}}^-$ has its minima.

%--------------------------------CONCLUSION----------------------------------------------
\section{Conclusion} \label{sec:Con}
We have explored the plane and stripe wave phases of a weakly interacting SOC BEC on a square lattice in the presence of a Zeeman field. It was found that while the helicity projection provides an excellent approximation for the PW phase with only one condensate momentum, it fails to describe the SW phase which has two condensate momenta. While the PW phase has a phonon minimum at its condensate momentum showing a nonzero, anisotropic speed of sound, the minima in the SW phase excitation spectrum show quadratic behavior and hence zero sound velocity. At strong SOC, the phase diagram based on the operator independent part of the Hamiltonian provides a good description of the system. The PW phase develops an energetic instability close to the PWSW transition line, while the SW phase becomes dynamically unstable when approaching the PWSW transition line.

%--------------------------------ACKNOWLEDGEMENTS-------------------------------------------
\section*{Acknowledgments}
We acknowledge useful discussions with Sigve Solli. 
We acknowledge funding from 
the Research Council of Norway Project No. 250985 ``Fundamentals of Low-Dissipative Topological Matter", and the Research Council of Norway through its Centres of Excellence funding scheme, Project No. 262633, ``QuSpin".

%---------------------------APPENDICES------------------------------------
\appendix

%-----------------------PW MATRIX SPIN
\section{PW Phase Matrix in Spin Basis} \label{sec:PWmatelem}
In the spin basis, the PW phase operator basis, $\boldsymbol{A}_{\boldsymbol{k}}$, is
\begin{equation*}
     (A_{\boldsymbol{k}}^{\uparrow},A_{2\boldsymbol{k}_{01}-\boldsymbol{k}}^{\uparrow}, A_{\boldsymbol{k}}^{\downarrow},A_{2\boldsymbol{k}_{01}-\boldsymbol{k}}^{\downarrow},A_{\boldsymbol{k}}^{\uparrow\dagger},A_{2\boldsymbol{k}_{01}-\boldsymbol{k}}^{\uparrow\dagger}, A_{\boldsymbol{k}}^{\downarrow\dagger},A_{2\boldsymbol{k}_{01}-\boldsymbol{k}}^{\downarrow\dagger})^T.
\end{equation*}
The matrix takes the form \eqref{eq:SWMat} with $M_1 = $
\begin{gather*}
    \begin{pmatrix}
    M_{11}(\boldsymbol{k}) & 0 & M_{13}(\boldsymbol{k}) & 0 \\
    0 & M_{11}(2\boldsymbol{k}_{01}-\boldsymbol{k}) & 0 & M_{13}(2\boldsymbol{k}_{01}-\boldsymbol{k}) \\
    M_{13}^{*}(\boldsymbol{k}) & 0 & M_{33}(\boldsymbol{k}) & 0 \\
    0 & M_{13}^{*}(2\boldsymbol{k}_{01}-\boldsymbol{k}) & 0 & M_{33}(2\boldsymbol{k}_{01}-\boldsymbol{k}) \\
    \end{pmatrix}
\end{gather*}
and
\begin{gather*}
    M_2^* = 
    \begin{pmatrix}
    0 & M_{52} & 0 & M_{72} \\
    M_{52} & 0 & M_{72} & 0 \\
    0 & M_{72} & 0 & M_{74} \\
    M_{72} & 0 & M_{74} & 0 \\
    \end{pmatrix}.
\end{gather*}
The matrix elements are
\begin{align}
    \begin{split}
    \label{eq:PWelem}
        M_{11}(\boldsymbol{k}) &= \epsilon_{\boldsymbol{k}} + Unx + G_{k_0}^\uparrow, \\
        G_{k_0}^\uparrow &= 4t\cos k_0  + \abs{s_{\boldsymbol{k}_{01}}}\sqrt{\frac{1-x}{x}},  \\
        M_{33}(\boldsymbol{k}) &= \epsilon_{\boldsymbol{k}} + Un(1-x) + G_{k_0}^\downarrow, \\
        G_{k_0}^\downarrow &=  4t\cos k_0  + \abs{s_{\boldsymbol{k}_{01}}}\sqrt{\frac{x}{1-x}}, \\
        M_{13}(\boldsymbol{k}) &= s_{\boldsymbol{k}} + Un\alpha\sqrt{x(1-x)} e^{i(\theta_{1}^{\downarrow}-\theta_{1}^{\uparrow})}, \\
        M_{52} &= Unx e^{i2\theta_{1}^{\uparrow}}, \\
        M_{72} &= Un\alpha\sqrt{x(1-x)} e^{i(\theta_{1}^{\uparrow} + \theta_{1}^{\downarrow})}, \\
        M_{74} &= Un(1-x) e^{i2\theta_{1}^{\downarrow}}.
    \end{split}
\end{align}
Where it leads to simplifications we have used the fact that $\theta_{1}^\downarrow - \theta_{1}^\uparrow = \pi/4$ minimizes the free energy in the PW phase.
%$\gamma_{\boldsymbol{k}_{01}}+\Delta\theta_1 = \pi$

%--------------------SW MATRIX SPIN-----------------------------------------------
\section{SW Phase Matrix in Spin Basis} \label{sec:SWmatelem}
Due to the form of the matrix $M_{\boldsymbol{k}}$ \eqref{eq:SWMat} together with the fact that $M_1$ is Hermitian and $M_2$ is symmetric \cite{Tsallis, Xiao} it is enough to specify rows $1,2,7$ and $8$ of $M_1$ and $M_2^*$. The rest of the matrix may then be filled, and any unspecified elements are $0$. With the values of the variational parameters found to minimize the free energy inserted, we have
\begin{equation}
\begin{aligned}
    M_{1, \textrm{row} 1} = (&M_{1,1}(\boldsymbol{k}), 0, M_{1,3}, 0, M_{1,3}^*, 0,\\
    &s_{\boldsymbol{k}}, 0, M_{1,9}, 0, -iM_{1,9}^*, 0), \\
    M_{1, \textrm{row} 2} = (&0, M_{1,1}(\boldsymbol{k}), 0, M_{1,3}, 0, M_{1,3}^*, \\
    &0, -s_{\boldsymbol{k}}, 0, M_{1,9}, 0, -iM_{1,9}^*), \\
    M_{1, \textrm{row} 7} = (&s_{\boldsymbol{k}}^*, 0, iM_{1,9}, 0, M_{1,9}^*, 0, \\
    &M_{7,7}(\boldsymbol{k}), 0, M_{7,9}, 0, M_{7,9}^*, 0), \\
    M_{1, \textrm{row} 8} = (&0, -s_{\boldsymbol{k}}^*, 0, iM_{1,9}, 0, M_{1,9}^*, \\
    &0, M_{7,7}(\boldsymbol{k}), 0, M_{7,9}, 0, M_{7,9}^*), 
\end{aligned} 
\end{equation}
\begin{equation}
\begin{aligned}
    M_{2, \textrm{row} 1}^* = (&0, M_{13,2}, 0, M_{13,4}, 0, M_{13,6}, \\
    &0, 0, 0, M_{13,10}, 0, M_{13,12}),\\
    M_{2, \textrm{row} 2}^* = (&M_{13,2}, 0, M_{13,4}, 0, M_{13,6}, 0,\\
    &0, 0, M_{13,10}, 0,M_{13,12}, 0), \\
    M_{2, \textrm{row} 7}^* = (&0, 0, 0, M_{13,10}, 0, M_{13,12}, \\
    &0, M_{19,8}, 0, M_{19,10}, 0, M_{19,12}),\\
    M_{2, \textrm{row} 8}^* = (&0, 0, M_{13,10}, 0, M_{13,12}, 0, \\
    &M_{19,8}, 0, M_{19,10}, 0, M_{19,12}, 0).
\end{aligned} 
\end{equation}
The matrix elements in $M_1$ are
\begin{align}
    \begin{split}
        \label{eq:SWelem}
        M_{1,1}(\boldsymbol{k}) &= \epsilon_{\boldsymbol{k}}+\frac{Un}{2}\big(x+(1-x)\alpha\big)+G_{k_0}^\uparrow, \\
        G_{k_0}^\uparrow &= 4t\cos k_0  + \abs{s_{\boldsymbol{k}_{01}}}\sqrt{\frac{1-x}{x}}, \\
        M_{7,7}(\boldsymbol{k}) &= \epsilon_{\boldsymbol{k}}+\frac{Un}{2}(1-x+x\alpha)+G_{k_0}^\downarrow, \\
        G_{k_0}^\downarrow &= 4t\cos k_0 + \abs{s_{\boldsymbol{k}_{01}}}\sqrt{\frac{x}{1-x}}, \\
        M_{1,3} &= \frac{Un}{4}e^{i(\theta_1^\uparrow-\theta_3^\uparrow)}\left(2x-(1-x)\alpha \right), \\
        M_{1,9} &= \frac{Un\alpha}{4}\sqrt{x(1-x)}e^{i(\theta_1^\downarrow-\theta_3^\uparrow)}, \\
        M_{7,9} &= \frac{Un}{4}e^{i(\theta_1^\uparrow-\theta_3^\uparrow)}\left(-2(1-x)+x\alpha \right) , \\
    \end{split}
\end{align}
while the elements in $M_2^*$ are
\begin{equation}
\begin{aligned}
M_{13,2} &= Unx e^{i(\theta_1^\uparrow+\theta_3^\uparrow)},  \\
M_{13,4} &= \frac{Un}{4}xe^{i2\theta_1^\uparrow}, \\
M_{13,6} &= \frac{Un}{4}xe^{i2\theta_3^\uparrow}, \\
 M_{13,10} &= \frac{Un\alpha}{4}\sqrt{x(1-x)}e^{i(\theta_1^\downarrow+\theta_1^\uparrow)},  \\
 M_{13,12} &= \frac{Un\alpha}{4}\sqrt{x(1-x)}e^{i(\theta_3^\downarrow+\theta_3^\uparrow)},  \\
M_{19,8} &= Un(1-x) e^{i(\theta_1^\downarrow+\theta_3^\downarrow)},  \\
M_{19,10} &= \frac{Un}{4}(1-x)e^{i2\theta_1^\downarrow},   \\
 M_{19,12} &= \frac{Un}{4}(1-x)e^{i2\theta_3^\downarrow}. \\
\end{aligned}
\end{equation}
The angles are left unspecified in the elements where inserting them would not lead to simplifications. We mentioned that setting a certain subset of these matrix elements to zero leads to a linear behavior close to the minima of the excitation spectrum. One choice is $M_{1,3} = M_{7,9} = M_{1,9} = M_{13,2} = M_{13,6} = M_{13,12} = M_{19,8} = M_{19,12} = 0$ with $\Omega = 0$. These are connected to the scatterings (c), (d), (f) and (g) in figure \ref{fig:SWscatterings}, i.e. mostly the scatterings involving both condensate momenta, supporting the claim that the quadratic behavior in the SW phase is caused by the presence of more than one condensate momentum.

\section{SW Phase Matrix in Helicity Projection} \label{sec:SWmatelemhel}
The operator vector is
\begin{align}
    \begin{split}
        \boldsymbol{C}_{\boldsymbol{k}} = (&C_{\boldsymbol{k}}, C_{-\boldsymbol{k}}, C_{\boldsymbol{p}_+}, C_{\boldsymbol{q}_+}, C_{\boldsymbol{p}_-}, C_{\boldsymbol{q}_-}, \\
        &C_{\boldsymbol{k}}^\dagger, C_{-\boldsymbol{k}}^\dagger, C_{\boldsymbol{p}_+}^\dagger, C_{\boldsymbol{q}_+}^\dagger, C_{\boldsymbol{p}_-}^\dagger, C_{\boldsymbol{q}_-}^\dagger)^T .
    \end{split}
\end{align}
The $12 \cross 12$ matrix $M_{\boldsymbol{k}}^H$ is of the form \eqref{eq:SWMat}. It is enough to specify rows $1$ and $2$ of $M_1^H$ and $M_2^{H*}$.
\begin{equation}
\begin{aligned}
M_{1, \textrm{row} 1}^H &= (M_{11}(\boldsymbol{k}), 0, M_{13}(\boldsymbol{k}), 0, M_{15}(\boldsymbol{k}), 0), \\
M_{1, \textrm{row} 2}^H &= (0, M_{11}(-\boldsymbol{k}), 0, M_{13}(-\boldsymbol{k}), 0, M_{15}(-\boldsymbol{k})), \\
M_{2, \textrm{row} 1}^{H*} &= (0, M_{72}(\boldsymbol{k}), 0, M_{74}(\boldsymbol{k}), 0, M_{76}(\boldsymbol{k})), \\ 
M_{2, \textrm{row} 2}^{H*} &= (M_{72}(\boldsymbol{k}), 0, M_{74}(-\boldsymbol{k}), 0, M_{76}(-\boldsymbol{k}), 0 ). \\ 
\end{aligned}
\end{equation}
The matrix elements are defined as follows
\begin{equation}
    \begin{aligned}
    M_{11}(\boldsymbol{k}) &= \lambda_{\boldsymbol{k}}^{-} -\lambda_{\boldsymbol{k}_{01}}^{-}\\
    &-\frac{Un}{2}(3u_{\boldsymbol{k}_{01}}^4+3\abs{v_{\boldsymbol{k}_{01}}}^4+2\alpha u_{\boldsymbol{k}_{01}}^2\abs{v_{\boldsymbol{k}_{01}}}^2) \\
    &+Un\big(2u_{\boldsymbol{k}_{01}}^2u_{\boldsymbol{k}}^2 + 2\abs{v_{\boldsymbol{k}_{01}}}^2\abs{v_{\boldsymbol{k}}}^2 \\
    &+ \alpha u_{\boldsymbol{k}_{01}}^2\abs{v_{\boldsymbol{k}}}^2 + \alpha \abs{v_{\boldsymbol{k}_{01}}}^2u_{\boldsymbol{k}}^2\big), \\
    M_{13}(\boldsymbol{k}) &=\frac{Un}{4}\big( 2u_{\boldsymbol{k}_{01}}^2u_{\boldsymbol{k}}u_{\boldsymbol{p}_+} -2 \abs{v_{\boldsymbol{k}_{01}}}^2 v_{\boldsymbol{k}}^* v_{\boldsymbol{p}_+}\\
    &-\alpha u_{\boldsymbol{k}_{01}}v_{\boldsymbol{k}_{01}}v_{\boldsymbol{k}}^*u_{\boldsymbol{p}_{+}} + \alpha u_{\boldsymbol{k}_{01}}^2 v_{\boldsymbol{k}}^* v_{\boldsymbol{p}_{+}}\\
    &+ \alpha v_{\boldsymbol{k}_{01}}^*u_{\boldsymbol{k}_{01}}u_{\boldsymbol{k}}v_{\boldsymbol{p}_{+}} - \alpha \abs{v_{\boldsymbol{k}_{01}}}^2u_{\boldsymbol{k}}u_{\boldsymbol{p}_{+}}\big) , \\
    M_{15}(\boldsymbol{k}) &= \frac{Un}{4}\big(  2u_{\boldsymbol{k}_{01}}^2u_{\boldsymbol{k}}u_{\boldsymbol{p}_-} -2 \abs{v_{\boldsymbol{k}_{01}}}^2 v_{\boldsymbol{k}}^* v_{\boldsymbol{p}_-}\\
    &+\alpha u_{\boldsymbol{k}_{01}}v_{\boldsymbol{k}_{01}}v_{\boldsymbol{k}}^*u_{\boldsymbol{p}_{-}}  + \alpha u_{\boldsymbol{k}_{01}}^2 v_{\boldsymbol{k}}^* v_{\boldsymbol{p}_{-}}\\
    &- \alpha v_{\boldsymbol{k}_{01}}^*u_{\boldsymbol{k}_{01}}u_{\boldsymbol{k}}v_{\boldsymbol{p}_{-}} - \alpha \abs{v_{\boldsymbol{k}_{01}}}^2u_{\boldsymbol{k}}u_{\boldsymbol{p}_{-}}\big), \\
    M_{72}(\boldsymbol{k}) &= Un\left(u_{\boldsymbol{k}_{01}}^2 u_{\boldsymbol{k}}^2 + (v_{\boldsymbol{k}_{01}}^*)^2 v_{\boldsymbol{k}}^2\right), \\
    M_{74}(\boldsymbol{k}) &= \frac{Un}{4}\big( u_{\boldsymbol{k}_{01}}^2u_{\boldsymbol{k}}u_{\boldsymbol{q}_+} + (v_{\boldsymbol{k}_{01}}^*)^2 v_{\boldsymbol{k}} v_{\boldsymbol{q}_+} \\
    & +\alpha u_{\boldsymbol{k}_{01}}v_{\boldsymbol{k}_{01}}^*v_{\boldsymbol{k}}u_{\boldsymbol{q}_{+}} + \alpha u_{\boldsymbol{k}_{01}}v_{\boldsymbol{k}_{01}}^*u_{\boldsymbol{k}}v_{\boldsymbol{q}_{+}}\big), \\
    M_{76}(\boldsymbol{k}) &= \frac{Un}{4}\big( u_{\boldsymbol{k}_{01}}^2u_{\boldsymbol{k}}u_{\boldsymbol{q}_-} + (v_{\boldsymbol{k}_{01}}^*)^2 v_{\boldsymbol{k}} v_{\boldsymbol{q}_-} \\
    & -\alpha u_{\boldsymbol{k}_{01}}v_{\boldsymbol{k}_{01}}^*v_{\boldsymbol{k}}u_{\boldsymbol{q}_{-}} - \alpha u_{\boldsymbol{k}_{01}}v_{\boldsymbol{k}_{01}}^*u_{\boldsymbol{k}}v_{\boldsymbol{q}_{-}}\big).
    \end{aligned}
\end{equation}
When $\Omega = 0$ and $\lambda_R/t < \sqrt{6}$, $M_{13}(\boldsymbol{k}) = 0$ and $M_{76}(\boldsymbol{k}) = 0$ around $\boldsymbol{k}_{01}$, while $M_{15}(\boldsymbol{k}) = 0$ and $M_{74}(\boldsymbol{k}) = 0$ around $-\boldsymbol{k}_{01}$. Such cancellations are considered erroneous upon comparison with the spin basis.

%---------------------------------------FREE ENERGY-------------------------------------------------------------------
\section{Free Energy} \label{sec:SWF}
In this appendix we will give an overview of the methods involved in calculating the free energy in the SW phase, and hence determining the values of the variational parameters. The use of commutators when setting up the matrix $M_{\boldsymbol{k}}$ gives a shift $-\sum_{\boldsymbol{k}}^{'} (M_{1,1}(\boldsymbol{k})+M_{7,7}(\boldsymbol{k}))/2$ of the operator independent part of the Hamiltonian. Employing the BV transformation, we numerically transform $H_1$ to the basis where $H_2$ is diagonal. The terms that are linear in excitation operators may then be removed by completing squares using terms from $H_2$. We shift some operators by complex numbers, which does not alter their interpretation since their commutation relations are conserved. Finally, this procedure leads to a shift of the free energy by a real number. More details can be found in \cite{KM}.

We consider the free energy at zero temperature, such that $F = \langle H \rangle$. Using that $ \langle B_{\boldsymbol{k},\sigma}^{\dagger}B_{\boldsymbol{k},\sigma}\rangle = 0$ for $\boldsymbol{k} \neq \pm \boldsymbol{k}_{01}$ we may now calculate the free energy numerically at a given set of parameters. To find the minimum of $F$ with respect to a variational parameter, we vary it while keeping the other variational parameters set to the values that minimize $H_0$. The result is that $k_0 = k_{0m}$ and $x$ equal to the value that minimizes $H_0$ also minimizes $F$ to a good approximation. Upon choosing $\theta_1^\uparrow$ as a free parameter, the angles that minimize $F$ are
\begin{equation}
\label{eq:SWangles}
    \theta_1^\downarrow = \theta_3^\uparrow = \theta_1^\uparrow + \frac{\pi}{4} \mbox{\quad and \quad} \theta_3^\downarrow = \theta_3^\uparrow + \frac{5\pi}{4}.
\end{equation}
A more rigorous approach would be to use simulated annealing \cite{simulatedannealing} to find the global minimum of the free energy in terms of the set of variational parameters. This was performed on the SW phase with no Zeeman field in \cite{Jonas}, and again the values that minimize $H_0$ were found to minimize $F$.

In general we find that the values of the variational parameters which minimize $F = \langle H \rangle$ can be approximated very well by the values that minimize $H_0$. This can be understood from the order of the terms in the Hamiltonian \eqref{eq:MFTH}. $H_0$ is of order $N^2/N_s$, $H_1$ of order $N\sqrt{N}/N_s$ and $H_2$ of order $N/N_s$. When $n = N/N_s = 1$ and $N_s >>1$ it is natural that $H_0$ dominates the minimization. In experiments, typical lattice sizes are $N_s \sim 1-3 \cdot 10^{5}$ while $n$ is most often of order unity \cite{expBEC2dfillingSpielman, expBEC3dfillingKetterle, PethickSmith}. We have therefore set $n=1$ when producing the figures, and have assumed that $\mu$ is set to the value which ensures this.

%--------------------------SPECIAL MOMENTA----------------------------------------------

\section{Special Momenta and Energetic Instability in SW Phase} \label{sec:special}
This appendix will briefly mention some subtle points not considered in the paper. Firstly, there are special momenta that require a separate treatment \cite{KM}. Considering the SW phase operator vector \eqref{eq:SWbra} at $\boldsymbol{k} = \boldsymbol{0}$ and $\pm2\boldsymbol{k}_{01}$ several elements become equal, which is not acceptable in the BV transformation. Additionally, at $\boldsymbol{k} = \pm 3 \boldsymbol{k}_{01}$ there are elements involving the condensate momenta $\pm\boldsymbol{k}_{01}$ in \eqref{eq:SWbra}. Such terms should have been excluded from the sum $\sum_{\boldsymbol{k}\boldsymbol{k}'}^{''}$ as mentioned after \eqref{eq:H2}.

However, the physical interpretation of these results is problematic. For instance, the special eigenvalues found at $\pm2\boldsymbol{k}_{01}$ do not correspond to the eigenvalues of $M_{\pm2\boldsymbol{k}_{01}}J$ suggesting the excitation spectrum is discontinuous. For $\alpha$ close to $3$ and $\Omega$ close to $0$ these special eigenvalues are lower than the minimum of the excitation spectrum, which seems to indicate energetic instabilities. However, we suggest treating this as an artifact of the BV diagonalization, rather than an indication of instability in the SW phase. On physical grounds we expect a continuous excitation spectrum.

For $\alpha = 3$ and no Zeeman field we find that the anomalous modes in the excitation spectrum are zero for all $\boldsymbol{k}$. Hence, the lowest energy band is zero for all $\boldsymbol{k}$, indicating that the phase is unstable. This seems to be caused by the fact that $|M_{1,3}|$ and $|M_{7,9}|$ become equal to $|M_{13,4}|$, $|M_{13,6}|, |M_{19,10}|$ and $|M_{19,12}|$. This means that the transition rates of the scatterings (c) and (d) become equal to the rates of (e) and (f) in figure \ref{fig:SWscatterings}. These matrix elements are related to scatterings where the $\boldsymbol{k}$-dependent particles have the same spin. A similar behavior is not found for $\Omega > 0$, since then $x \neq 1/2$. At least when ignoring any indications of energetic instability from the special momenta and keeping the SOC strength large, this energetic instability appears to be located at a single point in $\Omega-\alpha$-space and was therefore omitted in the paper.

%-------------------------------Linear behavior in interacting Bose gases-----------------------------------
\section{Linear Excitation Spectra in Interacting Bose Gases} \label{sec:linearquadratic}
In this appendix we review Bogoliubov's treatment of the weakly interacting Bose gas, focusing on the delicate cancellations that are required for the interactions to give a linear behavior  of the excitation spectrum close to its minimum. From this, we will give arguments for why the excitation spectrum in the PW phase showed a linear behavior in momentum, while the SW phase showed a quadratic behavior. We follow the treatments in \cite{PethickSmith, Pitaevskii, abrikosov} with the exception that we will consider the Bose gas defined on a 2D square lattice. The Hamiltonian is
\begin{equation}
    H = \sum_{\boldsymbol{k}} (\epsilon_{\boldsymbol{k}} -\mu) A_{\boldsymbol{k}}^\dagger A_{\boldsymbol{k}} + \frac{U}{2N_s}\sum_{\boldsymbol{k}\boldsymbol{k'}\boldsymbol{p}\boldsymbol{p'}} A_{\boldsymbol{k}}^\dagger A_{\boldsymbol{k'}}^\dagger A_{\boldsymbol{p}} A_{\boldsymbol{p'}} \delta_{\boldsymbol{k}+\boldsymbol{k'},\boldsymbol{p}+\boldsymbol{p'}}.
\end{equation}
It is assumed that the condensate is dominant, such that terms containing more than two excitation operators are neglected. Also, condensate operators are replaced by their expectation values. From now on we focus solely on the quadratic part of the Hamiltonian, $H_2$, which is
\begin{align}
    \begin{split}
    \label{eq:H2beforeBose}
        H_2 =  &\sum_{{\boldsymbol{k}} \neq {\boldsymbol{0}}} \left(\mathcal{E}_{\boldsymbol{k}} + Un \right) A_{\boldsymbol{k}}^\dagger A_{\boldsymbol{k}} \\
        &+ \frac{Un}{2}\sum_{{\boldsymbol{k}} \neq {\boldsymbol{0}}} \left( A_{\boldsymbol{k}}A_{-\boldsymbol{k}} + A_{\boldsymbol{k}}^\dagger A_{-\boldsymbol{k}}^\dagger \right).
    \end{split}
\end{align}
Here, we introduced $\mathcal{E}_{\boldsymbol{k}} = 4t+ \epsilon_{\boldsymbol{k}}$ which varies between $0$ and $8t$ and has a quadratic minimum at $\boldsymbol{k} = \boldsymbol{0}$.
Before applying the Bogoliubov transformation \cite{bogoliubov1947theory}, it is convenient to rewrite $H_2$ to
\begin{align}
    \begin{split}
    \label{eq:H2rewriteBose}
        H_2 = \frac{1}{2}\sum_{{\boldsymbol{k}} \neq {\boldsymbol{0}}}\bigg[ &\left(\mathcal{E}_{\boldsymbol{k}} + Un \right) \left(A_{\boldsymbol{k}}^\dagger A_{\boldsymbol{k}} +A_{-\boldsymbol{k}} A_{-\boldsymbol{k}}^\dagger  \right) \\
        &+ Un\left(A_{\boldsymbol{k}} A_{-\boldsymbol{k}} + A_{\boldsymbol{k}}^\dagger A_{-\boldsymbol{k}}^\dagger \right)\bigg].
    \end{split}
\end{align}
We refer to \cite{PethickSmith, Pitaevskii, abrikosov}. Notice that in \eqref{eq:H2beforeBose} the off-diagonal terms are $Un/2$ whereas the diagonal term contains $Un$. Then, we use commutators and let $\boldsymbol{k} \to -\boldsymbol{k}$. The off-diagonal operator combinations are symmetric under this combined operation. Hence, the factor of $1/2$ discrepancy is absent in \eqref{eq:H2rewriteBose}. Now, the term $Un$ appears both in the diagonal term $\mathcal{E}_{\boldsymbol{k}} + Un$ as well as the off-diagonal term. This leads to a cancellation of the constant term $(Un)^2$ in the expression $(\mathcal{E}_{\boldsymbol{k}} + Un)^2 - (Un)^2$ which is involved in obtaining the excitation spectrum $\omega_{\boldsymbol{k}} = \sqrt{\mathcal{E}_{\boldsymbol{k}}(\mathcal{E}_{\boldsymbol{k}}+2Un)}$. For $\omega_{\boldsymbol{k}}$ to show a linear behavior close to its minimum, it was essential that the $(Un)^2$ term canceled exactly.

We note that similar cancellations are found for two-component condensates at zero momentum \cite{LS, KM}. These give linear minima in the excitation spectra which are obtained using the BV transformation. Now, we move on to the two phases studied in this paper, and focus on the spin basis. 

In the PW phase, we encounter similar effects as in the interacting Bose gas considered above. The PW phase matrix elements are presented in appendix \ref{sec:PWmatelem}. Notice that $|M_{52}|$ is contained in $M_{11}(\boldsymbol{k})$, $|M_{74}|$ is contained in $M_{33}(\boldsymbol{k})$ and $|M_{72}|$ is contained in $M_{13}(\boldsymbol{k})$. While we are unable to get analytic expressions for the excitation spectrum in this case, it is likely these relations between the matrix elements allow the cancellations necessary to get a linear spectrum. These relations between the matrix elements are caused by the way we rewrite the Hamiltonian by using commutators and, in this case, symmetrizing around $\boldsymbol{k}_{01}$. The interaction terms in the $M_2$ matrix elements originally have a factor of $1/2$ difference from the interaction terms in the $M_1$ matrix elements. However, since the operator products associated with these terms are symmetric under the combined operation of using commutators and letting $\boldsymbol{k} \to 2\boldsymbol{k}_{01}-\boldsymbol{k}$ this discrepancy disappears, analogously to what happened in the interacting Bose gas. A key aspect the PW phase shares with the case considered in detail above, is that the condensate exists only at one momentum, albeit finite instead of zero. 

In the SW phase, the condensate exists at two distinct finite momenta. This is a qualitatively new aspect not shared by the PW phase or the standard case with condensate at zero momentum. Moreover, it is clearly a lattice effect.    
In the SW phase, all the scatterings represented in figure \ref{fig:SWscatterings} are relevant for the quadratic part of the Hamiltonian. In the interacting Bose gas, there are scatterings like (a), (b) and (g), if we let the condensate momenta go to zero. In the PW phase, only (a) and (e) are relevant. In these cases it is the symmetries in (g) or (e) which lead to a linear behavior. Scattering (e) is symmetric about $\boldsymbol{k}_{01}$, scattering (f) is symmetric about $\boldsymbol{k}_{03}$, while scattering (g) is symmetric about $\boldsymbol{k} = \boldsymbol{0}$. 

%It is a unique feature of condensates with more than one condensate momenta that we can have a simultaneous presence of several scatterings with different symmetries.

%As usual, we use commutators in rewriting the Hamiltonian in the SW phase. Additionally, we need to symmetrize, so that the matrix takes the form \eqref{eq:SWMat} where $M_1$ is Hermitian and $M_2$ is symmetric. This is necessary for the theory of the BV transformation \cite{Tsallis, Xiao} to be applicable. 

As usual, we rewrite the sum over $\boldsymbol{k}$ in the Hamiltonian in the SW phase, using commutators and a symmetry.  
Choosing the symmetry of scattering (g), i.e. symmetrizing about $\boldsymbol{k} = \boldsymbol{0}$ by making $-\boldsymbol{k}$-terms explicit, is the only choice which does not make the operator basis larger. The resulting matrix elements are represented in appendix \ref{sec:SWmatelem}.

Because the scattering (g) can be constructed in two ways due to the two condensate momenta, its corresponding matrix elements get an extra factor of $2$. Therefore, e.g. $|M_{13,2}| = Unx$ is a factor of $2$ larger than the corresponding $U$-dependent term in $M_{1,1}(\boldsymbol{k})$. The scatterings (e) and (f) are not symmetric about $\boldsymbol{k} = \boldsymbol{0}$. Hence, their matrix elements retain their factor of $1/2$ discrepancy from the diagonal elements, also after the symmetrization of the Hamiltonian. Therefore, e.g. $|M_{13,4}| = Unx/4$ is a factor of $1/2$ smaller than the corresponding $U$-dependent term in $M_{1,1}(\boldsymbol{k})$.

Though we do not have analytic expressions for the excitation spectrum, we believe these discrepancies are the reason for the quadratic behavior in the SW phase. Essentially, the conditions for delicate cancellations which give linear behavior are not present. 
%We have connected this to the presence of two condensate momenta in the following way.
The presence of two condensate momenta yields a greater set of scattering processes, which gives a larger operator basis, and a larger matrix from which the excitation spectrum is obtained. This makes the eigenvalue problem more complicated, and reduces the likelihood of cancellations. In particular, the simultaneous presence of the scatterings (e), (f) and (g) with separate symmetries is a unique feature of more than one condensate momentum, and seems to be essential for why the necessary cancellations are not present to give a linear behavior.

%----------------------------BIBLIOGRAPHY---------------------------------------
\bibliography{main.bbl}

\end{document}